%% file: acl_latex.tex
\documentclass[11pt]{article}

\usepackage[final]{acl}

\usepackage{times}
\usepackage{latexsym}
\input{math_commands.tex}

\usepackage{hyperref}
\usepackage{url}
\usepackage{enumitem}
\usepackage{threeparttable}
\usepackage{makecell}
\usepackage{booktabs}
\usepackage{amssymb}
\usepackage{pifont}
\usepackage{multirow}
\usepackage{array}
\usepackage{tikz}
\usepackage{xcolor}

\usepackage{wrapfig}
\usepackage{amsmath}
\usepackage[T1]{fontenc}
\usepackage[utf8]{inputenc}
\usepackage{microtype}
\usepackage{inconsolata}
\usepackage{graphicx}
\usepackage{algorithm}
\usepackage{algorithmic}
\usepackage{tabularx}

\title{From Past To Path: Masked History Learning for\\
Next-Item Prediction in Generative Recommendation}

\author{
  \textbf{Kaiwen Wei\textsuperscript{1}\thanks{These authors contributed equally to this work.}},
  \textbf{Kejun He\textsuperscript{1}\footnotemark[1]},
  \textbf{Xiaomian Kang\textsuperscript{2}},
  \textbf{Jie Zhang\textsuperscript{3}},
  \textbf{Yuming Yang\textsuperscript{1}}, \\
  \textbf{Li Jin\textsuperscript{4}},
  \textbf{Zhenyang Li\textsuperscript{5}},
  \textbf{Jiang Zhong\textsuperscript{1}\thanks{Corresponding authors.}},
  \textbf{He Bai\textsuperscript{3}},
  \textbf{Junnan Zhu\textsuperscript{2}\footnotemark[2]} \\
  \textsuperscript{1}College of Computer Science, Chongqing University \\
  \textsuperscript{2}MAIS, Institute of Automation, Chinese Academy of Sciences \\
  \textsuperscript{3}Independent Researcher \\
  \textsuperscript{4}Aerospace Information Research Institute, Chinese Academy of Sciences \\
  \textsuperscript{5}Department of Computer Science, City University of Hong Kong \\
  \texttt{\href{mailto:weikaiwen@cqu.edu.cn}{weikaiwen@cqu.edu.cn}, 
  \href{mailto:hekejun@stu.cqu.edu.cn}{hekejun@stu.cqu.edu.cn},
  \href{mailto:junnan.zhu@nlpr.ia.ac.cn}{junnan.zhu@nlpr.ia.ac.cn}}
}

\begin{document}

\maketitle


\begin{abstract}
Generative recommendation, which directly generates item identifiers, has emerged as a promising paradigm for recommendation systems. However, this left-to-right paradigm inherently biases the model towards local contexts, failing to capture deeper historical dependencies necessary for understanding complex user intents.
To address this limitation, we propose \textbf{Masked History Learning (MHL)}, a novel training framework that shifts the objective from simple next-step prediction to deep comprehension of history. MHL augments the standard autoregressive objective with an auxiliary task of reconstructing masked historical items, compelling the model to understand ``why'' an item path is formed from the user's past behaviors, rather than just ``what'' item comes next.
We introduce two key contributions to enhance this framework: (1) an \textbf{entropy-guided masking} policy that intelligently targets the most informative historical items for reconstruction, and (2) a \textbf{curriculum learning} scheduler that progressively transitions from history reconstruction to future prediction.
Experiments on three public datasets show that our method significantly outperforms state-of-the-art generative models, highlighting that a comprehensive understanding of the past is crucial for accurately predicting a user's future path. 
The code is available at \url{https://github.com/CQU-MM-Intelligent-Lab/MHL}.
\end{abstract}

\section{Introduction}

Recommender systems have become essential tools for navigating the vast digital landscape, evolving from collaborative filtering~\citep{wang2015collaborative, li2024surveydeepneuralnetworks, chen2018survey} to sequential models that capture user behavior dynamics~\citep{purificato2024usermodelinguserprofiling, yuan2023where, he2023survey}. A new paradigm, \textit{generative recommendation}~\citep{rajput2023recommender, muennighoff2025generative}, has recently emerged, offering powerful new ways to model user preferences. This approach adapts pre-trained language models like T5~\citep{rajput2023recommender, bao2023tallrec} and utilizes large language models~\citep{hou2025generative} to directly generate a sequence of semantic IDs representing the items to be recommended~\citep{hua2023how, zhai2024actions}, thus providing unprecedented flexibility.

\begin{figure}[t] 
 \centering 
 \includegraphics[width=0.8 \columnwidth]{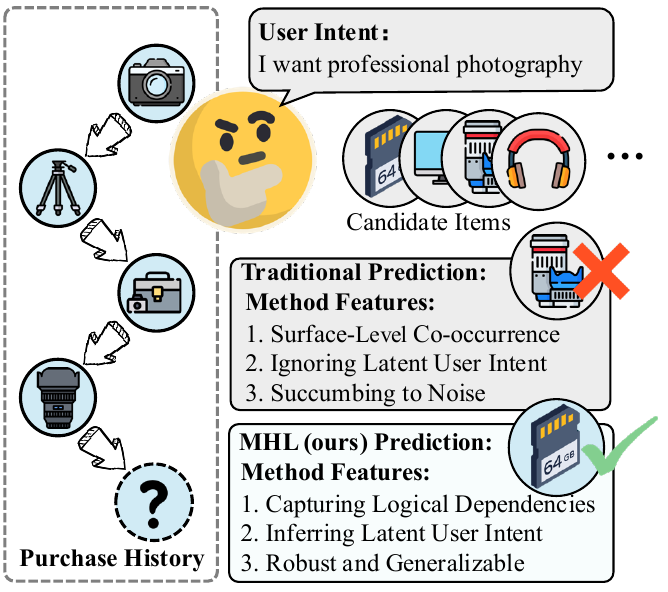} 
 \caption{Comparison between the traditional generative recommendation system and our MHL framework.} 
 \label{fig:intro} 
\end{figure}


However, despite their architectural diversity, these models share a fundamental limitation: they are trained almost exclusively to predict the next single item, rather than to understand the path that led there. This narrow focus on autoregressive next-item prediction, while intuitive, prioritizes local transitions over global understanding of user behavior. We contend that this paradigm yields models proficient in local forecasting yet deficient in global comprehension, rendering them susceptible to noise and short-term deviations (i.e., short-term myopia). Our preliminary study (see Figure~\ref{fig:pilot_test-set-length} in Appendix~\ref{appdix:pilot}) reveals a startling phenomenon, i.e., existing SOTA models suffer a \textbf{performance collapse} when recent interaction history is truncated, while MHL maintains robustness and \textbf{outperforms the leading baseline by over 40\%}. This exposes their over-reliance on recency bias rather than deep intent understanding. In contrast, MHL maintains robustness by leveraging global context.

For example, as shown in Figure~\ref{fig:intro}, consider the purchasing path of a photography enthusiast, who interacts with the following items in order: \textit{camera body}, \textit{tripod}, \textit{camera bag}, and \textit{camera lens}. Although the ground truth for the subsequent purchase is the \textit{memory card}, existing models, fixated on recent item (\textit{camera lens}), often incorrectly predict other lens-related accessories. The user's intention is a direct continuation of purchasing the initial \textit{camera body}, but this intention is obscured by intermediate items. Due to the inability to fully internalize the underlying intention associations behind items along the purchasing path, existing autoregressive models are trained merely to predict ``what comes next," but cannot effectively understand ``why this path matters."

To address this limitation, we introduce \textbf{Masked History Learning} (MHL), a novel training framework for generative recommendation. Specifically, we augment next-item prediction with an auxiliary objective of reconstructing masked items within historical paths. This approach shifts the learning paradigm from predicting results to understanding the process, yielding three key advantages: 
\textbf{(a) Capturing Logical Dependencies.} MHL compels the model to understand the intrinsic associations between masked items and other items, thereby shifting the focus from statistical co-occurrence to the logical structure of a user's path. For example, by reconstructing the masked historical item ``tripod", the model is forced to learn that ``tripod" is a logical complement to a ``camera body" purchase.
\textbf{(b) Inferring Latent User Intent.} The deep understanding of paths enables models to look beyond a user's explicit behaviors and infer the latent intent driving them. Models can learn to comprehend a coherent and higher-level goal (like ``pursuing professional photography") from seemingly disparate items (such as cameras, bags, and future accessories). 
\textbf{(c) Learning Robust and Generalizable Representations.} The history reconstruction objective inherently enhances the quality of the learned representations. To reconstruct history accurately, the model must prioritize strong, logically consistent signals while learning to discount irrelevant or noisy interactions that provide poor contextual clues. This focus on the core signal results in item representations that are more stable and less susceptible to incidental deviations in behavior.

We validate the proposed MHL at multiple granularities: item-, token-, and mixed-level, consistently observing performance gains. To refine this learning process, we introduce two key innovations. First, moving beyond random masking, we propose an adaptive strategy guided by information theory~\citep{mackay2002information}. We selectively mask items sharing high \textbf{entropy} with others, creating challenging training signals that focus on   significant behavioral patterns. Second, we employ \textbf{curriculum learning}~\citep{bengio2009curriculum} to connect the history reconstruction training with autoregressive inference. The training process begins with a warmup phase~\citep{he2016deep} using random masking, followed by a transition to a high masking ratio guided by entropy to build deep contextual understanding. The ratio is  gradually reduced to prepare the model for path generation.

We conduct extensive experiments on three categories of the Amazon Reviews 2014 dataset~\citep{mcauley2015image}. The results demonstrate that understanding the past significantly enhances the model's ability to predict future paths, outperforming state-of-the-art baselines on metrics like Recall@K and NDCG@K. The contributions of this paper can be summarized as follows:

\begin{itemize}[leftmargin=10pt]
    \item We identify a key limitation in generative recommenders: training dominated by next-step prediction overlooks deep understanding of user history. We address this by proposing \textit{Masked History Learning}, which learns to reconstruct a user's \textit{past} to better predict their future \textit{path}.
    \item We design two strategies to enhance our framework: \textit{entropy-guided masking} to focus on the most informative historical parts, and \textit{curriculum learning} to bridge the gap between understanding history and generating future paths.
    \item Extensive experiments on three categories of the Amazon Reviews 2014 dataset validate our approach's effectiveness, achieving new state-of-the-art results for generative recommendation.
\end{itemize}

\section{Related Work}
\textbf{Sequential Recommendation.} Sequential recommendation models user behavior over time to predict future interactions. Early methods use Markov chains to capture item-to-item transitions~\citep{rendle2010factorizing}. Deep learning has since transformed this field. Modern approaches employ various neural architectures including recurrent neural networks~\citep{hidasi2016session, li2017neural, yue2024linear}, convolutional neural networks~\citep{tang2018personalized}, Transformers~\citep{kang2018self, sun2019bert4rec}, and graph neural networks~\citep{chang2021sequential, wu2019session}. 

While most sequential models are trained autoregressively, some studies have explored alternative learning objectives that go beyond simple next-item prediction. BERT4Rec~\citep{sun2019bert4rec} and S$^3$-Rec~\citep{zhou2020s3rec} use masked item prediction with bidirectional encoders to learn rich contextual representations for discriminative recommendation. These models randomly mask items in user sequences and learn to reconstruct them using full bidirectional context. This approach helps models capture richer dependencies compared to purely left-to-right training. Despite this success, the generative recommendation paradigm has remained largely autoregressive.

\paragraph{Generative Recommendation.} Recent advances in generative models have promoted the transformation of recommendation systems from discriminative to generative~\citep{rajput2023recommender}. Inspired by generative retrieval~\citep{tay2022transformer, wang2022neural}, these methods tokenize items into discrete semantic identifiers. Sequence-to-sequence models can then directly generate these identifiers as recommendations.

Two main approaches have emerged in generative recommendation. The first leverages Large Language Models (LLMs) through zero-shot prompting~\citep{gao2023chatrec, jesse2023leveraging} and instruction tuning~\citep{muennighoff2025generative} to align LLMs with user behaviors. The second focuses on semantic ID-based generation, where items are first encoded as discrete token sequences~\citep{rajput2023recommender} derived from quantizing dense representations~\citep{hua2023how, wang2024learnable}, then autoregressively decoded to produce recommendations~\citep{zhai2024actions,lin2025spacetime}.

Despite their flexibility and scalability, existing generative recommenders~\citep{rajput2023recommender, wang2024learnable, hou2025generative} share a common limitation: they rely almost exclusively on autoregressive training that predicts the next item token given previous tokens. This left-to-right approach focuses on local transitions but may miss internal dependencies and underlying user intent.

\paragraph{Our Contribution.} Our study addresses this gap by introducing history reconstruction learning to generative recommendation. Unlike BERT4Rec and S$^3$-Rec, which employ masked prediction within bidirectional encoders to learn representations for discriminative scoring tasks, our proposed MHL augments standard \textit{unidirectional, decoder-only} model with an auxiliary historical reconstruction objective. This design preserves the model's native autoregressive generation capability while enriching the training signal through deeper historical understanding. We further introduce entropy-guided masking to focus learning on the most informative historical patterns and curriculum learning to seamlessly transition from history understanding to future path generation. Together, these contributions establish a new training paradigm for generative recommenders that emphasizes understanding the past to better predict future paths.

\begin{figure*}[t]
  \centering
  \includegraphics[width=\textwidth]{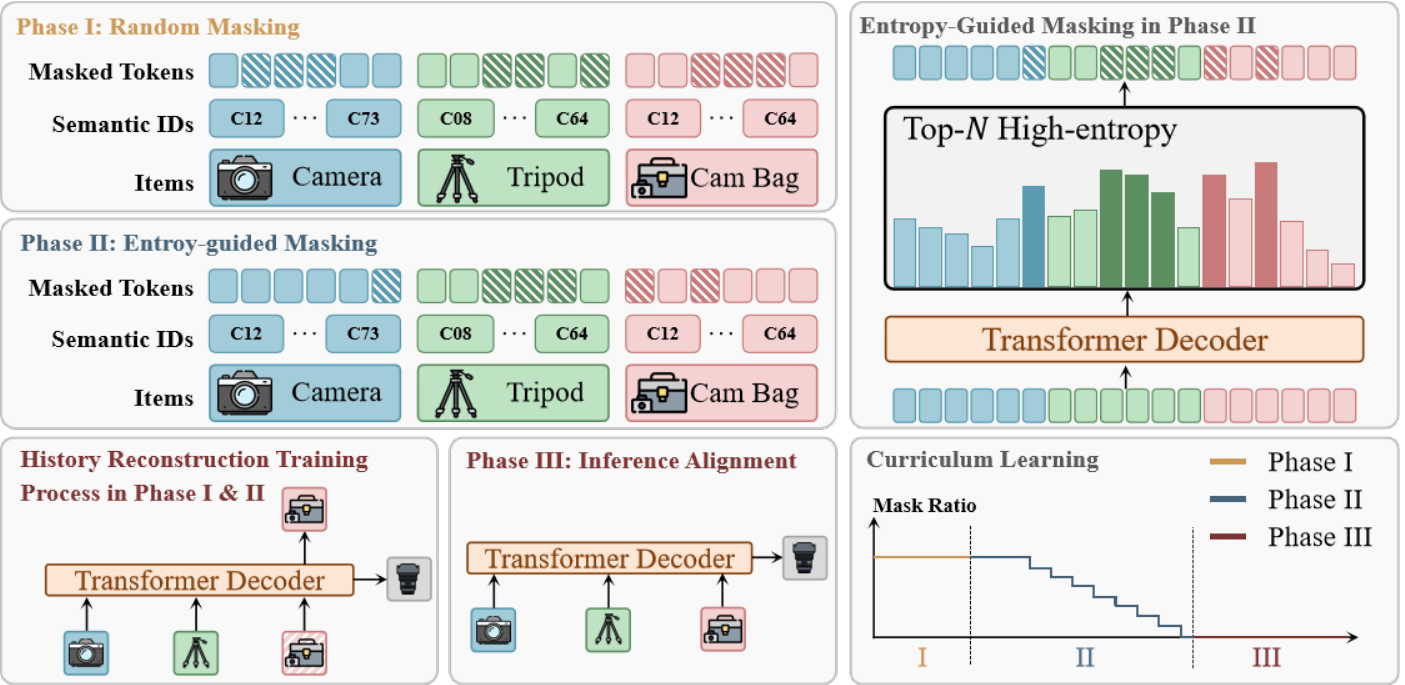}
  \caption{MHL framework overview. Online entropy computation selects challenging positions for masking. Curriculum training progresses from random to adaptive masking, then aligns with inference conditions.}
  \label{fig:framework}
\end{figure*}

\section{Method}
We propose \textbf{Masked History Learning (MHL)}, a training framework that augments generative recommendation with an auxiliary objective of reconstructing masked historical items. As illustrated in Figure~\ref{fig:framework}, MHL jointly optimizes next-item prediction and masked history reconstruction, guided by entropy-based masking and curriculum scheduling.

\subsection{Preliminaries}
\label{sec:preliminary}

\paragraph{Semantic ID Representation.}
Recent generative recommendation systems~\citep{rajput2023recommender, hou2025generating} represent each item as a sequence of discrete semantic tokens. In this work, each item $i$ from the item set $\mathcal{I}$ is encoded as a $K$-digit semantic ID $\phi(i)=\{w_i^1,\dots,w_i^K\}$, where each digit $w_i^k$ is drawn from a codebook $\mathcal{W}^k$. We construct a unified vocabulary $\mathcal{V}=\mathcal{V}_{\text{code}}\cup\mathcal{V}_{\text{mask}}$, where codebook tokens occupy non-overlapping ID ranges and position-aware mask tokens enable unambiguous reconstruction (details in Appendix~\ref{app:vocab}).

\paragraph{Base Architecture.}
Given a user history $S_T=(i_1,\dots,i_T)$, item $i_t$ obtains its item-level representations $H_t$ by performing mean pooling on $K$-digit embeddings, and generates context hidden state $\mathbf{h}_t \in \mathbb{R}^d$ using a Transformer decoder. For prediction, $K$ lightweight heads $\{g_k\}_{k=1}^K$ derive digit-wise states $\mathbf{h}_t^k=g_k(\mathbf{h}_t)$, which are classified using temperature-scaled cosine similarity over codebook embeddings. We denote the resulting probability distribution over codebook $\mathcal{W}^k$ as $P_\theta^k(\cdot \mid \mathbf{h}_t)$. Full architectural details are provided in Appendix~\ref{app:arch}.

\subsection{Masked History Learning}
\label{sec:mhl}

The core idea of MHL is to train the model not only to predict \textit{what comes next}, but also to understand \textit{why the historical path is formed}. We achieve this by jointly optimizing two objectives: (i) next-item prediction and (ii) masked history reconstruction.

\paragraph{Next-Item Prediction.}
Given the final hidden state $\mathbf{h}_T$, we predict the semantic ID of the next item $i_{T+1}$ using $K$ parallel codebook classifiers:
\begin{equation}
\small
\mathcal{L}_{\text{next}}=-\frac{1}{K}\sum_{k=1}^K \log P_\theta^k(w_{i_{T+1}}^k\mid \mathbf{h}_T).
\end{equation}

\paragraph{Masked History Reconstruction.}
We mask selected positions in the user history and train the model to reconstruct the original tokens. Let $\mathcal{M}$ denote the set of masked positions. For each masked token $w_{i_t}^k \in \mathcal{M}$, we predict it from the contextualized state at position $t$:
\begin{equation}
\small
\mathcal{L}_{\text{mask}}=-\frac{1}{|\mathcal{M}|}\sum_{w_{i_t}^k\in\mathcal{M}} \log P_\theta^k(w_{i_t}^k\mid \mathbf{h}_t).
\end{equation}
This auxiliary objective forces the model to capture logical dependencies between items---not merely statistical co-occurrence, but the underlying intent that connects them.

\paragraph{Multi-Granularity Masking Strategies.}
We design three masking granularities to provide diverse learning signals:

\begin{itemize}[leftmargin=12pt, itemsep=2pt]
    \item \textbf{Item-level masking} replaces all $K$ digits of selected items. This treats each item as a complete semantic unit, compelling the model to learn inter-item dependencies that align with user-level behaviors.
    
    \item \textbf{Token-level masking} replaces individual digits within items. This enables fine-grained learning of intra-item token relationships, yielding more compact representations and better generalization by preventing over-reliance on complete item patterns.
    
    \item \textbf{Mixed-level masking} stochastically applies item- or token-level masking per item, balancing coarse and fine granularity for robustness.
\end{itemize}

\paragraph{Overall Objective.}
The final training objective combines both losses:
\begin{equation}
\small
\mathcal{L}_{\text{MHL}} = \lambda_1 \mathcal{L}_{\text{next}} + \lambda_2 \mathcal{L}_{\text{mask}},
\end{equation}
where $\lambda_1$ and $\lambda_2$ are hyperparameters balancing next-item generation and history reconstruction.




\subsection{Entropy-Guided Masking}
\label{sec:entropy_mask}

Standard random masking treats all historical interactions equally, ignoring the fact that user behaviors have uneven information density. To address this, we propose \textbf{Entropy-Guided Masking}, an uncertainty-driven strategy inspired by salience-aware methods in language generation~\citep{liu2025sara} that dynamically targets the most ambiguous and informative positions for reconstruction.

\paragraph{Online Entropy Computation.}
We employ an \textbf{online self-assessment} mechanism to estimate the uncertainty of each historical item. In each training step, we perform a gradient-free forward pass on the unmasked sequence $S_T$ using the current model parameters $\theta$. For each position $t$ and codebook $k$, the predictive uncertainty is quantified by the Shannon entropy of the output distribution:
\begin{equation}
\small
\mathcal{H}_t^k = -\sum_{v \in \mathcal{W}^k} P_\theta^k(v \mid \mathbf{h}_t) \log P_\theta^k(v \mid \mathbf{h}_t).
\end{equation}
where $\mathbf{h}_t$ is the hidden state encoding the history up to $t$. High entropy $\mathcal{H}_t^k$ indicates that the model is uncertain about the item at $t$ based on the context. These high-entropy positions often correspond to \textbf{critical decision points} or \textbf{complex semantic units}, such as the \textit{tripod} in Figure~\ref{fig:intro}, which is a logical complement to the camera body rather than a generic frequent item.

\paragraph{Adaptive Mask Selection.}
Based on the computed uncertainty, we adaptively construct the mask set $\mathcal{M}$ using a stochastic budgeting strategy. 
First, we aggregate token-level entropies into an item-level importance score $\bar{\mathcal{H}}_t = \frac{1}{K}\sum_{k=1}^K \mathcal{H}_t^k$. 
Then, we rank all historical positions in descending order of their importance scores.
To prevent the model from overfitting to a fixed masking density, we sample the mask budget $N$ from a uniform distribution $\mathcal{U}(1, \lfloor \gamma \cdot T \rfloor)$, where $\gamma$ is the current curriculum ratio.
Finally, we select the top-$N$ positions with the highest entropy to form $\mathcal{M}$.
This adaptive strategy ensures that the reconstruction objective ($\mathcal{L}_{\text{mask}}$) is always applied to the positions where the model lacks understanding, thereby maximizing the gradient information gain and forcing the model to capture deeper historical dependencies.


\subsection{Curriculum Training Scheduler}
\label{sec:curriculum}

Directly applying entropy-guided masking from the start leads to unstable training, as the model's early-stage entropy estimates are unreliable. Moreover, training with masking creates a discrepancy with inference, where no masking occurs. We address both issues through a three-phase curriculum building upon adaptive data selection strategies~\citep{commonIT,rao2025dynamicsamplingadaptsiterative}:

\paragraph{Phase I: Random Masking Warm-up.}
Training begins with random masking at a low ratio. This phase serves two purposes: (1) establishing stable optimization before entropy estimates become reliable, and (2) building baseline reconstruction ability through simple, unbiased masking.

\paragraph{Phase II: Entropy-Guided Masking with Adaptive Decay.}
Once the model stabilizes, we switch to entropy-guided masking to focus on the most informative positions. The masking ratio $\gamma$ starts high to encourage deep historical understanding. When validation performance plateaus, we decay $\gamma$ exponentially:
\begin{equation}
\small
\gamma \leftarrow \max(\gamma_{\min},\ \gamma \cdot \eta),
\end{equation}
where $\gamma_{\min}$ is the minimum ratio threshold and $\eta \in (0,1)$ is the decay factor. This gradually shifts the model's focus from history reconstruction toward future prediction.

\paragraph{Phase III: Inference Alignment.}
Finally, we set $\gamma=0$ and train exclusively with $\mathcal{L}_{\text{next}}$. This phase eliminates the train-test discrepancy: since inference involves no masking, inference-aligned training ensures the model is fully aligned with its deployment condition. This smooth transition---from reconstruction to generation---bridges the gap between understanding ``why this path matters'' and predicting ``what comes next.''

The complete training procedure is summarized in Algorithm~\ref{alg:curriculum} (Appendix~\ref{app:algorithm}).

\begin{table*}[t!]
\centering
\resizebox{\textwidth}{!}{
\begin{tabular}{>{\centering\arraybackslash}p{2.0cm} cccc cccc cccc}
\toprule
\multirow{2}{*}{\makecell[c]{\rule{0pt}{2.5ex}\textbf{Model}}}
& \multicolumn{4}{c}{\textbf{Beauty}}
& \multicolumn{4}{c}{\textbf{Toys and Games}}
& \multicolumn{4}{c}{\textbf{Sports and Outdoors}} \\
\cmidrule(lr){2-5} \cmidrule(lr){6-9} \cmidrule(lr){10-13}
 & R@5 & N@5 & R@10 & N@10
 & R@5 & N@5 & R@10 & N@10
 & R@5 & N@5 & R@10 & N@10 \\
\midrule
\multicolumn{13}{c}{\textbf{Item ID-based}} \\
\midrule
\multirow{7}{*}{\parbox[t]{2.0cm}{\raggedright Caser\\GRU4Rec\\HGN\\BERT4Rec\\SASRec\\FDSA\\S$^3$-Rec}}
 & .0205 & .0131 & .0347 & .0176 & .0166 & .0107 & .0270 & .0141 & .0116 & .0072 & .0194 & .0097 \\
 & .0164 & .0099 & .0283 & .0137 & .0097 & .0059 & .0176 & .0084 & .0129 & .0086 & .0204 & .0110 \\
 & .0325 & .0206 & .0512 & .0266 & .0321 & .0221 & .0497 & .0277 & .0189 & .0120 & .0313 & .0159 \\
 & .0203 & .0124 & .0347 & .0170 & .0116 & .0071 & .0203 & .0099 & .0115 & .0075 & .0191 & .0099 \\
 & .0387 & .0249 & .0605 & .0318 & .0463 & .0306 & .0675 & .0374 & .0233 & .0154 & .0350 & .0192 \\
 & .0267 & .0163 & .0407 & .0208 & .0228 & .0140 & .0381 & .0189 & .0182 & .0122 & .0288 & .0156 \\
 & .0387 & .0244 & .0647 & .0327 & .0443 & .0294 & .0700 & .0376 & .0251 & .0161 & .0385 & .0204 \\
\midrule
\multicolumn{13}{c}{\textbf{Semantic ID-based}} \\
\midrule
\multirow{5}{*}{\parbox[t]{2.0cm}{\raggedright RecJPQ\\VQ-Rec\\TIGER\\HSTU\\RPG*}}
 & .0311 & .0167 & .0482 & .0222 & .0331 & .0182 & .0484 & .0231 & .0141 & .0076 & .0220 & .0102 \\
 & .0457 & .0317 & .0664 & .0383 & .0497 & .0346 & .0737 & .0423 & .0208 & .0144 & .0300 & .0173 \\
 & .0454 & .0321 & .0648 & .0384 & .0521 & .0371 & .0712 & .0432 & .0264 & .0181 & .0400 & .0225 \\
 & .0469 & .0314 & .0704 & .0389 & .0433 & .0281 & .0669 & .0357 & .0258 & .0165 & .0414 & .0215 \\
 & \underline{.0500} & \underline{.0358} & \underline{.0745} & \underline{.0436} & \underline{.0550} & \underline{.0386} & \underline{.0778} & \underline{.0460} & \underline{.0284} & \underline{.0197} & \underline{.0436} & \underline{.0246} \\
\midrule
\multirow{1}{*}{\parbox[t]{2.0cm}{\raggedright MHL (ours)}}
 & \textbf{.0574} & \textbf{.0424} & \textbf{.0795} & \textbf{.0495} & \textbf{.0672} & \textbf{.0489} & \textbf{.0903} & \textbf{.0564} & \textbf{.0359} & \textbf{.0249} & \textbf{.0511} & \textbf{.0298} \\
\bottomrule
\end{tabular}%
}
\caption{Performance comparison of Item ID-based and Semantic ID-based models across three datasets. * denotes results using \textbf{Sentence-T5-base} embeddings for fair comparison with our setup. \textbf{Best} and second-best results are bolded and \underline{underlined}.}
\label{tab:all_models_transposed}
\end{table*}

\section{Experiment}
\subsection{Experimental Settings}
\textbf{Dataset.} 
We evaluate models on three Amazon product categories: \textit{Sports and Outdoors}, \textit{Beauty}, and \textit{Toys and Games} from the Amazon Reviews 2014 dataset~\citep{mcauley2015image}. We preprocess each category with core-5 filtering~\citep{he2016ups}. This retains only users and items with at least five interactions to ensure sufficient density for sequential modeling. For item metadata, we concatenate title, price, brand, feature, categories, and description into natural language sentences. This facilitates semantic representation learning following recent practice in generative recommendation~\citep{wang2024learnable}. Table~\ref{tab:dataset_stats} from Appendix~\ref{appendix:statistic} shows detailed dataset statistics.

\paragraph{Baselines.}
We evaluate against comprehensive baselines in two categories: item ID-based methods and semantic ID-based approaches. Item ID-based methods operate directly on item IDs: {GRU4Rec}~\citep{hidasi2016session}, {HGN}~\citep{ma2019hierarchical}, {SASRec}~\citep{kang2018self}, 
FDSA~\citep{hao2023feature}, {BERT4Rec}~\citep{sun2019bert4rec}, {Caser}~\citep{tang2018personalized}, {S$^3$-Rec}~\citep{zhou2020s3rec}. Semantic ID-based approaches tokenize items into discrete semantic identifiers for generative recommendation: {VQRec}~\citep{hou2023learning}, {RecJPQ}~\citep{petrov2024recjpq}, {TIGER}~\citep{rajput2023recommender}, {HSTU}~\citep{zhai2024actions}, {RPG}~\citep{hou2025generating}.

\begin{table*}[ht]
\centering
\scriptsize
\setlength{\tabcolsep}{3pt}
\resizebox{\textwidth}{!}{%
\begin{tabular}{>{\centering\arraybackslash}p{1.5cm} >{\centering\arraybackslash}p{2.0cm} cccc cccc cccc}
\toprule
\multirow{2}{*}{\makecell[c]{\rule{0pt}{2.5ex}Mask\\Strategy}} & \multirow{2}{*}{\makecell[c]{\rule{0pt}{2.5ex}Curriculum\\Strategy}} 
 & \multicolumn{4}{c}{Beauty} 
 & \multicolumn{4}{c}{Toys and Games} 
 & \multicolumn{4}{c}{Sports and Outdoors} \\
\cmidrule(lr){3-6} \cmidrule(lr){7-10} \cmidrule(lr){11-14}
 &  & R@5 & N@5 & R@10 & N@10 
    & R@5 & N@5 & R@10 & N@10 
    & R@5 & N@5 & R@10 & N@10 \\
\midrule
\parbox[t]{1.5cm}{\raggedright No Mask} & \raggedright Direct Inference & .0500 & .0358 & .0745 & .0436 & .0550 & .0386 & .0778 & .0460 & .0284 & .0197 & .0436 & .0246 \\
\midrule
\multirow{5}{*}{\parbox[t]{1.5cm}{\raggedright Token-level}}
 & \raggedright Random                          & \underline{.0554} & .0379 & \textbf{.0797} & .0457 & \textbf{.0647} & .0402 & \textbf{.0914} & .0488 & .0321 & .0179 & \underline{.0482} & .0231 \\
 & \raggedright Entropy-guided                  & .0491 & .0338 & .0715 & .0410 & .0232 & .0151 & .0359 & .0191 & .0110 & .0070 & .0171 & .0090 \\
 & \raggedright R$\rightarrow$Inf               & \underline{.0554} & \underline{.0404} & .0791 & \underline{.0480} & .0623 & \underline{.0452} & .0876 & \textbf{.0533} & \textbf{.0337} & \underline{.0233} & \textbf{.0486} & \underline{.0281} \\
 & \raggedright E$\rightarrow$Inf               & .0518 & .0374 & .0736 & .0445 & .0471 & .0322 & .0702 & .0396 & .0264 & .0180 & .0402 & .0224 \\
 & \raggedright R$\rightarrow$E$\rightarrow$Inf & \textbf{.0574} & \textbf{.0424} & \underline{.0795} & \textbf{.0495} & \underline{.0644} & \textbf{.0456} & \underline{.0883} & \textbf{.0533} & \underline{.0334} & \textbf{.0239} & .0474 & \textbf{.0284} \\
 
\midrule
\multirow{5}{*}{\parbox[t]{1.5cm}{\raggedright Item-level}}
 & \raggedright Random                          & \underline{.0500} & .0355 & .0694 & .0418 & \underline{.0564} & \underline{.0392} & \underline{.0808} & \underline{.0470} & \textbf{.0287} & \underline{.0198} & \textbf{.0420} & \underline{.0240} \\
 & \raggedright Entropy-guided                  & .0441 & .0315 & .0660 & .0386 & .0522 & .0359 & .0750 & .0432 & .0225 & .0151 & .0336 & .0187 \\
 & \raggedright R$\rightarrow$Inf               & .0494 & \underline{.0361} & \underline{.0695} & \underline{.0426} & \textbf{.0575} & \textbf{.0410} & \textbf{.0816} & \textbf{.0487} & .0270 & .0188 & .0415 & .0234 \\
 & \raggedright E$\rightarrow$Inf               & .0467 & .0335 & .0672 & .0401 & .0545 & .0376 & .0768 & .0448 & \underline{.0286} & \textbf{.0199} & .0414 & \underline{.0240} \\
 & \raggedright R$\rightarrow$E$\rightarrow$Inf & \textbf{.0501} & \textbf{.0363} & \textbf{.0704} & \textbf{.0429} & .0553 & .0390 & .0802 & \underline{.0470} & .0282 & .0197 & \underline{.0419} & \textbf{.0241} \\
 
\midrule
\multirow{5}{*}{\parbox[t]{1.5cm}{\raggedright Mixed-level}}
 & \raggedright Random                          & .0521 & .0359 & \textbf{.0789} & .0444 & .0589 & .0393 & \textbf{.0859} & .0480 & \underline{.0318} & .0197 & \textbf{.0493} & \underline{.0254} \\
 & \raggedright Entropy-guided                  & .0491 & .0344 & .0728 & .0420 & .0566 & .0377 & .0833 & .0463 & .0285 & .0186 & .0428 & .0232 \\
 & \raggedright R$\rightarrow$Inf               & \textbf{.0542} & \textbf{.0391} & .0752 & \textbf{.0459} & \textbf{.0593} & \underline{.0420} & .0835 & \underline{.0498} & .0295 & \underline{.0203} & .0435 & .0248 \\
 & \raggedright E$\rightarrow$Inf               & .0520 & .0371 & .0743 & .0442 & .0535 & .0376 & .0761 & .0449 & .0289 & .0202 & .0418 & .0243 \\
 & \raggedright R$\rightarrow$E$\rightarrow$Inf & \underline{.0537} & \underline{.0384} & \underline{.0757} & \underline{.0455} & \underline{.0592} & \textbf{.0421} & \underline{.0845} & \textbf{.0503} & \textbf{.0325} & \textbf{.0228} & \underline{.0478} & \textbf{.0278} \\
\bottomrule
\end{tabular}%
}
\caption{Ablation study comparing masking strategies and curriculum learning approaches with codebook size 16 and mask ratio 0.15. \textbf{Best} and second-best results are bolded and \underline{underlined}.}
\label{tab:ablation}
\end{table*}

\paragraph{Evaluation Metrics.}
We evaluate recommendation performance using Recall@K and NDCG@K with K=5 and 10. Following prior works~\citep{kang2018self, rajput2023recommender, sun2019bert4rec, hou2025generating}, we adopt standard leave-one-out strategy. For each user sequence, the last item is reserved for testing, the second-to-last for validation, and the remaining items for training.

\paragraph{Implementation Details.} We encode item metadata (e.g., title, brand, price) with Sentence-T5-base~\citep{Muresan2022SentenceT5}. The resulting 768-dimensional embeddings are reduced to 128 dimensions via PCA and then discretized into sequences of 32 semantic tokens using FAISS-based optimized product quantization (OPQ)~\citep{ge2013optimized}. Our backbone is a Transformer decoder, identical to the one used in RPG~\citep{hou2025generating}, featuring a hidden size of 448, two layers, and four attention heads. The model is trained to jointly optimize next-item prediction and masked token reconstruction with equal weights. We apply an entropy-guided curriculum masking strategy, and early stopping is used when the mask ratio decays to zero. During optimization, we use AdamW with a learning rate of 5e-4, a batch size of 64, and cosine scheduling with 10k warmup steps. Inference is performed with graph-constrained beam search~\citep{hou2025generating} (beam size 50, 3 propagation steps). The models are trained for up to 300 epochs on NVIDIA RTX A6000 GPUs. More details can be found in Appendix~\ref{implemental_Details}.

\subsection{Experiment Results}
\textbf{Overall  Performance.} In all experiments, we performed three runs with different random seeds and report the average results to account for variability. Table~\ref{tab:all_models_transposed} presents the results across three Amazon product categories. We can find that MHL consistently achieves state-of-the-art performance. In addition, the results confirm that semantic ID-based models outperform traditional item ID-based approaches, with MHL leading all baselines, including strong competitors like TIGER and HSTU. The performance improvements are substantial. For example, MHL achieves a 37.6\% improvement over TIGER in the NDCG@5 score for \textit{Sports and Outdoors}. This validates our claim: understanding why a user path is formed is crucial for predicting what comes next. MHL's superior performance demonstrates three key benefits. First, by reconstructing masked historical items, the model learns logical dependencies between items rather than co-occurrence patterns. Second, the entropy-guided masking forces the model to focus on the most informative and challenging positions in user history, precisely where latent intent is obscured. Third, the curriculum learning bridges the gap between history understanding and future prediction, ensuring a smooth transition from learning ``why this path matters" to predicting ``what comes next". These targeted learning mechanisms enable MHL to consistently outperform baselines. The framework's effectiveness is particularly evident on the complex dataset like \textit{Sports and Outdoors}, where logical item relationships are more nuanced and user intent is harder to infer.

\begin{table*}[t]
\centering
\scriptsize
\setlength{\tabcolsep}{2.5pt}
\resizebox{\textwidth}{!}{%
\begin{tabular}{>{\centering\arraybackslash}p{2.5cm} >{\centering\arraybackslash}p{2.0cm} ccccc ccccc ccccc}
\toprule
\multirow{2}{*}{\makecell[c]{\rule{0pt}{2.5ex}Mask\\Strategy}} & 
\multirow{2}{*}{\makecell[c]{\rule{0pt}{2.5ex}Training\\Method}} 
 & \multicolumn{5}{c}{Beauty} 
 & \multicolumn{5}{c}{Toys and Games} 
 & \multicolumn{5}{c}{Sports and Outdoors} \\
\cmidrule(lr){3-7} \cmidrule(lr){8-12} \cmidrule(lr){13-17}
 &  & R@5 & N@5 & R@10 & N@10 & $\Delta$\%
    & R@5 & N@5 & R@10 & N@10 & $\Delta$\%
    & R@5 & N@5 & R@10 & N@10 & $\Delta$\% \\
\midrule
\parbox[t]{2.5cm}{\raggedright Text w/o Mask} & \raggedright RPG
 & .0297 & .0212 & .0439 & .0258 & --
 & .0323 & .0234 & .0446 & .0273 & --
 & .0134 & .0094 & .0203 & .0117 & -- \\
\midrule
\parbox[t]{2.5cm}{\raggedright MHL (ours)} & \raggedright R$\rightarrow$E$\rightarrow$Inf
 & .0338 & .0238 & .0483 & .0285 & \textbf{+10.5}
 & .0347 & .0249 & .0498 & .0297 & \textbf{+8.8}
 & .0150 & .0106 & .0237 & .0134 & \textbf{+14.5} \\
\bottomrule
\end{tabular}
}
\caption{Generalization study comparing MHL with RPG baseline on text token sequences using mixed masking. MHL uses mask ratio 0.15 and reconstruction loss weight 0.5. $\Delta$\% denotes the relative improvement of N@10.}
\label{tab:text_based_full}
\end{table*}

\begin{figure}[t] 
    \centering 
    \includegraphics[width=\columnwidth]{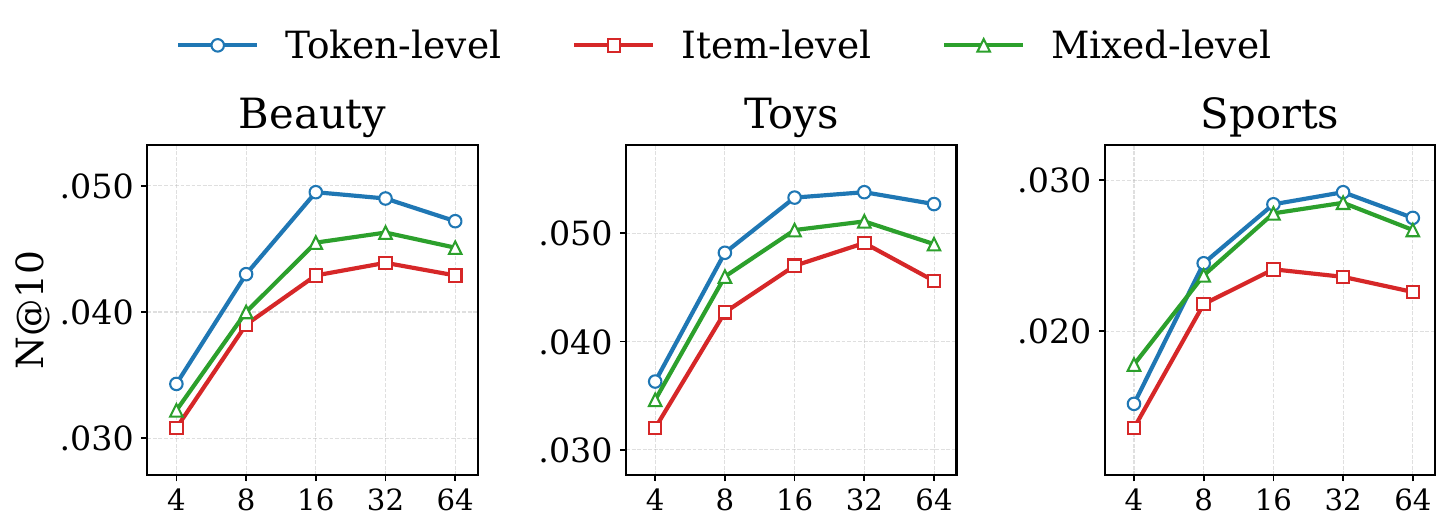} 
    \caption{Impact of varying codebook sizes on model performance across three mask strategies (Token-level, Item-level, Mixed-level) on 3 datasets.} 
    \label{fig:ablation_codebook_size} 
\end{figure}

\paragraph{Ablation Study.}  
We conduct systematic ablation studies to understand each component's contribution within MHL. We evaluate six model variants across three masking strategies: Direct Inference (Inf) without masking, Random masking (R), and Entropy-guided masking (E). We also test three curriculum learning strategies: R$\to$Inf, E$\to$Inf, and the complete R$\to$E$\to$Inf framework.
Table~\ref{tab:ablation} validates our framework design through three key findings. 
First, all masking variants significantly outperform direct inference, demonstrating that reconstructing user history provides a richer learning signal. Second, entropy-guided masking consistently surpasses random masking, indicating that targeting high-entropy predictions is more effective for guiding the model to understand user intent. Finally, the complete R$\to$E$\to$Inf curriculum learning framework achieves optimal performance. 

This validates our curriculum design: starting with basic pattern learning through random masking, progressing to targeted understanding via entropy guidance, and finally inference-only optimization. This progression mirrors the learning objective of transitioning from ``why this path matters" to ``what comes next".

\subsection{Further Analysis}

\paragraph{Impact of Semantic ID Length.}


We investigate the impact of codebook size (4–64) on model performance. As shown in Figure~\ref{fig:ablation_codebook_size}, performance first improves and then declines with increasing size: it rises as the size increases from 4 to 32, with sizes 16 and 32 yielding optimal results across datasets. Performance degrades at size 4 due to insufficient semantic granularity and slightly declines at size 64 because of sparsity issues. This trend reveals a fundamental trade-off between semantic expressiveness and statistical reliability. Excessively small codebooks (e.g., size 4) collapse distinct items into overly broad semantic clusters, impairing the model's ability to capture fine-grained preferences. Conversely, overly large codebooks (e.g., size 64) fragment the semantic space, leading to sparse code distributions where many semantic IDs receive insufficient training signals. The optimal sizes (16–32) strike a balance, allowing semantic IDs to capture meaningful item relationships while maintaining adequate sample density for robust parameter estimation. Based on these findings, we adopt a size of 16 for the Beauty dataset and 32 for Toys and Sports. Detailed results are provided in the Appendix~\ref{appdix:Detailed Performance Results Across Codebook Sizes}.


\begin{figure}[t] 
    \centering
    \includegraphics[width=\columnwidth]{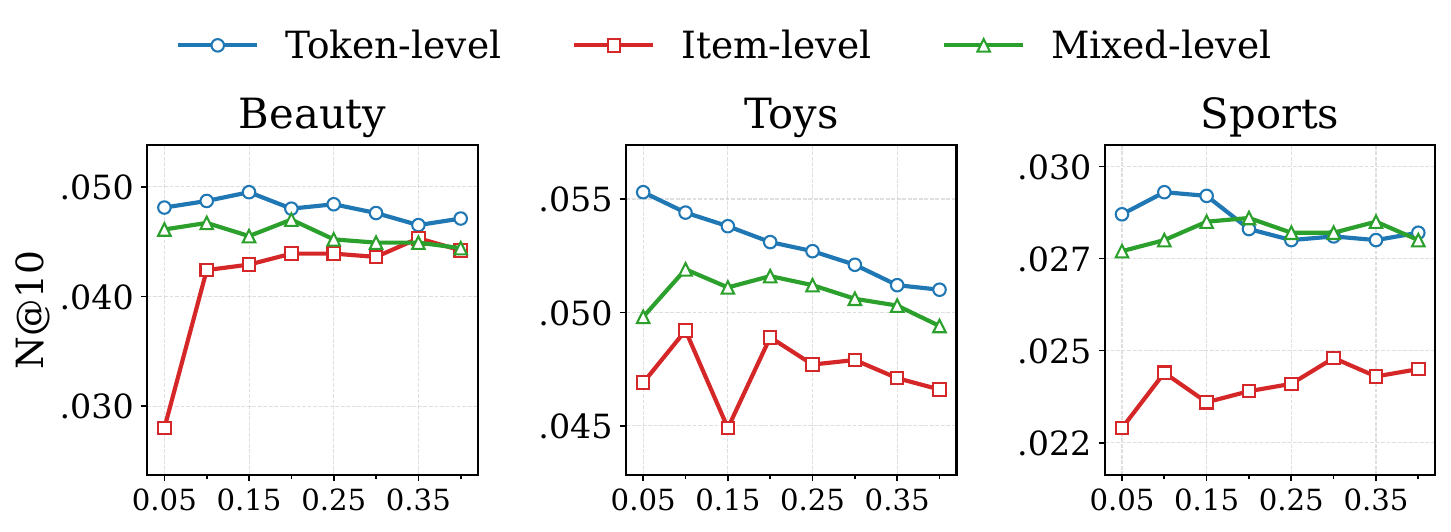} 
    \caption{Effect of different mask ratios on model performance under three mask strategies (Token-level, Item-level, Mixed-level) across 3 datasets.}
    \label{fig:ablation_mask_ratio} 
\end{figure}

\paragraph{Sensitivity to Masking Ratio.}


We vary the masking ratio from 0.05 to 0.40 to investigate how the reconstruction task difficulty affects model performance. As shown in Figure~\ref{fig:ablation_mask_ratio}, the masking ratio fundamentally balances contextual preservation and discriminative learning: lower ratios preserve richer historical signals for learning but may reduce the reconstruction objective's discriminative power, while higher ratios force stronger pattern extraction yet risk destroying critical user intent signals. MHL demonstrates strong robustness to these variations, with token-level masking achieving optimal performance at lower ratios (0.05--0.15) due to its finer granularity that enables precise targeting of uncertain predictions while preserving sequence context. Mixed-level masking provides a more balanced solution within the 0.10--0.20 range, exhibiting greater stability across ratios because its probabilistic combination of item and token masking creates natural curriculum effects where harder item-level reconstruction complements easier token-level recovery. Detailed results are in Appendix~\ref{appdix:Mask Ratio Sensitivity Analysis}.

\paragraph{Effect of Reconstruction Loss Weight.}  

\begin{figure}[t] 
    \centering 
    \includegraphics[width=\columnwidth]{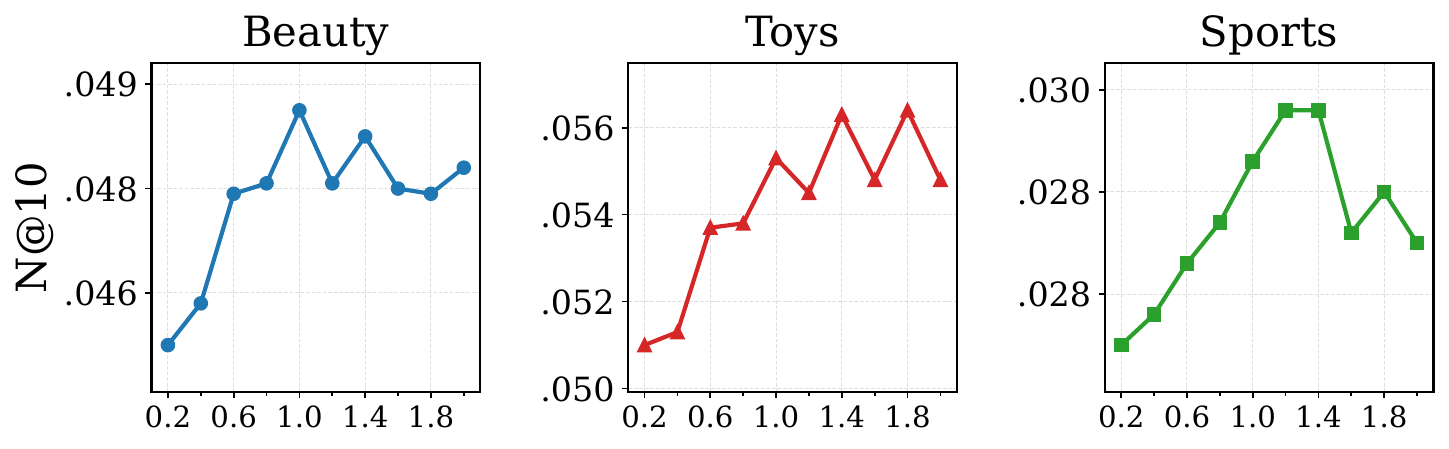} 
    \caption{Evaluation of model performance under different reconstruction loss values across multiple datasets, with a focus on N@10 performance.}
    \label{fig:ablation_recon_loss} 
\end{figure}

We vary the reconstruction loss weight from 0.2 to 2.0 while keeping the prediction loss fixed at 1.0 to examine how the auxiliary signal strength affects learning dynamics. As shown in Figure~\ref{fig:ablation_recon_loss}, performance degrades when the weight drops below 0.8, indicating that the influence of the reconstruction task weakens and diminishes the model's ability to learn from historical context. Conversely, excessive weight (above 2.0) skews the training objective toward reconstruction and compromises the model's capacity to learn prediction patterns during masking, ultimately sacrificing predictive discrimination. The optimal weights consistently fall within 1.0--1.8, striking a balance where reconstruction guides representation learning without overwhelming the primary prediction objective. Within this range, the auxiliary signal remains strong enough to leverage historical context during masking, yet sufficiently restrained to maintain the model's focus on future prediction. Detailed results are in Appendix~\ref{appdix:Recon Loss Sensitivity Analysis}.

\begin{figure}[t] 
    \centering
    \includegraphics[width=\columnwidth]{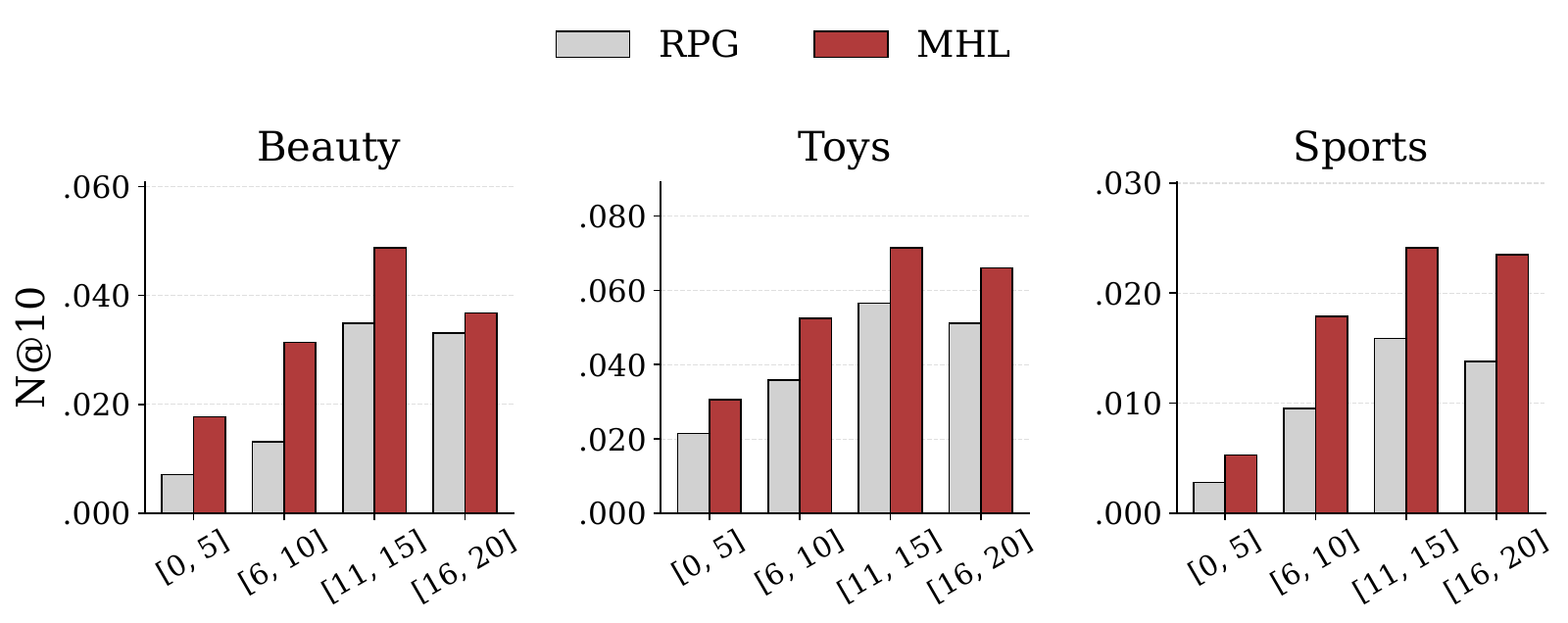} 
    \caption{Cold-start recommendation performance comparison between MHL and RPG across target item frequency bins on three datasets.}
    \label{fig:cold_start_comparison} 
\end{figure}


\paragraph{Generalizability to Text Sequences.}
To demonstrate MHL's broad applicability, we apply our framework to unstructured token sequences derived from item titles. Table~\ref{tab:text_based_full} compares our complete R$\to$E$\to$Inf strategy against the RPG baseline on raw text tokens.
MHL consistently outperforms RPG across all metrics on the Beauty dataset. This result is significant as it shows that MHL's core principle—reconstructing the past to predict the future—generalizes beyond semantic IDs to complex and noisy text sequences. The success on text sequences validates that MHL captures fundamental learning dynamics rather than exploiting specific properties of semantic ID representations.
This generalizability highlights MHL's potential for broader sequential modeling applications where historical context understanding is key for future prediction.


\paragraph{Cold-Start Recommendation.}


A practical challenge in recommendation systems is the cold-start problem, where target items have limited historical interactions. To evaluate MHL's robustness, we partition the test set by the training frequency of ground-truth target items into four groups: [0, 5], [6, 10], [11, 15], and [16, 20] occurrences, comparing MHL against RPG across all bins on three datasets. As shown in Figure~\ref{fig:cold_start_comparison}, both models exhibit a consistent inverted-U pattern: performance improves from [0, 5] to [11, 15], peaking at moderately frequent items, then declines at [16, 20]. This trend suggests that extremely rare items lack sufficient training signals while highly frequent items may suffer from popularity bias, whereas moderately frequent items strike an optimal balance. Importantly, MHL consistently outperforms RPG across all bins, with the gap being particularly significant for the coldest items, validating that history reconstruction enables the model to learn more robust representations from limited data and providing stronger generalization for infrequent items compared to pure next-item prediction. This superior cold-start performance demonstrates the strong ability of the masked history learning objective to capture deep semantic relationships between items, even when individual items have sparse interaction histories in the training data. Detailed results are in Appendix~\ref{appdix:Cold-Start Recommendation Analysis}.

\begin{table}[t!] 
\centering 
\footnotesize \setlength{\tabcolsep}{3pt} 
\begin{tabular}{c|c} 
\toprule 
\multicolumn{2}{c}{\textbf{Historical Purchase Sequence}} \\ 
\midrule 
\multicolumn{2}{p{0.95\linewidth}}{\centering Footwear Adhesive $\rightarrow$ Running Waist Pack $\rightarrow$ Cardio Trampoline $\rightarrow$ Heavyweight T-Shirt $\rightarrow$ BMX Pads $\rightarrow$ ?} \\ 
\midrule 
\textbf{MHL (Top-5)} & \textbf{RPG (Top-5)} \\ 
\midrule 
Youth Multi-Sport Helmet \textcolor{green}{\ding{51}} & Crew Sock \\ 
NBA Street Basketball & Eco Open Bottom Pant \\ 
Mini Basketball Hoop & Training T-shirt \\ 
Indoor/Outdoor Basketball & Jersey Pants \\ 
NBA Game Ball Mini & Long Sleeve Cotton T-Shirt \\ 
\bottomrule 
\end{tabular} 
\caption{Case study comparison between MHL and the RPG baseline. Top-5 recommendations are listed in descending order of predicted relevance.} 
\label{tab:case_study} 
\end{table}

\paragraph{Case Study.}
As illustrated in Table~\ref{tab:case_study}, the baseline RPG model, trained solely on autoregressive next-item prediction, often misinterprets a user's intent by overemphasizing transient, noisy signals, such as the mid-sequence clothing items. For example, its predictions for items like ``Crew Sock'' deviate from the user's primary and recurring interest in athletic gear and accessories. 
In contrast, MHL framework requires the model to reconstruct a user's historical trajectory, and encourages it to identify and prioritize the core underlying intent. MHL can look beyond short-term deviations and accurately predict the next item ``Youth Multi-Sport Helmet'', which aligns logically with the user's sustained interest in sports and protective equipment. 

\newcommand{\correctmark}{\tikz{\fill[green!70!black] (0,0) -- (0.05,0.15) -- (0.2,0); \draw[green!70!black, line width=0.5pt] (0.05,0.15) -- (0.15,-0.05);}}
\definecolor{darkgreen}{HTML}{006400}

\section{Conclusion}
Existing generative recommenders focus on predicting ``what comes next" but fail to understand ``why this path matters". We introduce MHL, a simple and effective framework that learns from masked history reconstruction alongside next-item prediction.
MHL incorporates entropy-guided masking to target informative historical positions and curriculum learning to transition from history understanding to future prediction. Experiments on three datasets show state-of-the-art performance and successful generalization to text-based Item IDs.
Our findings confirm that understanding the past is crucial for predicting the future. MHL represents a significant step toward recommendation systems that comprehend user behavior patterns rather than merely statistical co-occurrence.

\section*{Limitations}

MHL constructs semantic IDs solely from textual item metadata and does not explicitly incorporate visual or other multi-modal signals. Nevertheless, the proposed masking and reconstruction mechanism operates on discrete tokens and is modality-agnostic, making it applicable to multi-modal features once encoded into codebook-based representations; here we focus on text-only inputs to isolate the effect of masked history learning.

Entropy-guided masking introduces additional training-time cost due to an extra forward pass for uncertainty estimation. While this increases training time, it provides more informative supervision and leads to consistent performance gains, and the overhead is confined to training without affecting inference efficiency.

\section*{Ethics Statement}
This paper uses publicly available pretrained models and datasets. The datasets employed in this work are widely adopted in the research community and contain no private user data or personally identifiable information. All models are evaluated as-is, without any additional training or fine-tuning that could amplify harmful behaviors. We therefore believe that this study complies with the ACL Ethics Policy.

\section*{Acknowledgements}
We gratefully acknowledge the support of the National Key R\&D Program of China (No. 2024YFC2707805), the National Natural Science Foundation of China (No. 62176029, No. 62506050), the Open Competition Program of Chongqing Municipal Commission of Economy and Information Technology (No. YJX-202500100200X), the China Postdoctoral Science Foundation Funded Project (No. 2024M763867), and the Chongqing Higher Education Teaching Reform Research Project  (No. 242009).

The experimental and computational work in this research runs on the Huawei Cloud AI Compute Service. We appreciate the stable compute supply from this platform. We sincerely thank the anonymous reviewers for their insightful comments and constructive suggestions.

\bibliography{custom}

\appendix

\section{Vocabulary Construction Details}
\label{app:vocab}

Let $\mathcal{I}$ denote the item set. Each item $i \in \mathcal{I}$ is represented by a $K$-digit semantic ID using $K$ codebooks $\{\mathcal{W}^1, \dots, \mathcal{W}^K\}$, where $|\mathcal{W}^k| = V_k$ denotes the vocabulary size of the $k$-th codebook:
\begin{equation}
\phi(i) = \{w_i^1, \dots, w_i^K\}, \quad w_i^k \in \mathcal{W}^k.
\end{equation}

We construct a unified vocabulary that contains both semantic tokens and mask tokens:
\begin{equation}
\mathcal{V} = \mathcal{V}_{\text{code}} \cup \mathcal{V}_{\text{mask}},
\end{equation}
where $\mathcal{V}_{\text{code}} = \bigcup_{k=1}^K \mathcal{V}_{\text{code}}^k$ and each codebook block $\mathcal{V}_{\text{code}}^k$ occupies a non-overlapping ID range:
\begin{equation}
\mathcal{V}_{\text{code}}^k = \Big[ 1 + \sum_{j<k} V_j,\ \sum_{j \le k} V_j \Big].
\end{equation}

We allocate position-aware mask tokens $\mathcal{V}_{\text{mask}}$ for each position-digit pair $(t, k)$ up to the maximum sequence length $T_{\max}$:
\begin{equation}
\mathcal{V}_{\text{mask}} = \Big[ \sum_{k=1}^K V_k + 1,\ \sum_{k=1}^K V_k + K \cdot T_{\max} \Big],
\end{equation}
ensuring that each masked position can be uniquely identified, making the masking process lossless and unambiguous.

\section{Model Architecture Details}
\label{app:arch}

\paragraph{Token Embedding.}
Each token ID in $\mathcal{V}$ is mapped to a $d$-dimensional embedding by a shared lookup table (word token embedding, WTE). For a user history $S_T = (i_1, \dots, i_T)$, we expand each item into its $K$ semantic tokens $\{w_{i_t}^k\}_{k=1}^K$ and retrieve their embeddings:
\begin{equation}
e_t^k = \mathrm{WTE}(w_{i_t}^k) \in \mathbb{R}^d.
\end{equation}

\paragraph{Item-level Representation via Mean Pooling.}
We aggregate each item's $K$ digit embeddings into a single item-level representation:
\begin{equation}
H_t = \frac{1}{K} \sum_{k=1}^K e_t^k \in \mathbb{R}^d.
\end{equation}

\paragraph{Sequence Modeling.}
The item representations $\{H_1, \dots, H_T\}$ are fed into a Transformer decoder to obtain contextualized hidden states:
\begin{equation}
\mathbf{h}_{1:T} = \mathrm{Dec}(H_{1:T}), \quad \mathbf{h}_t \in \mathbb{R}^d.
\end{equation}

\paragraph{Digit-wise Prediction Heads.}
To predict each semantic digit independently, we use $K$ lightweight prediction heads $\{g_k\}_{k=1}^K$ that derive digit-specific states from the shared item state:
\begin{equation}
\mathbf{h}_t^k = g_k(\mathbf{h}_t), \quad k = 1, \dots, K, \quad \mathbf{h}_t^k \in \mathbb{R}^d.
\end{equation}

\paragraph{Cosine Classifier with Temperature Scaling.}
For codebook $k$, let $E^k \in \mathbb{R}^{V_k \times d}$ denote the embedding matrix where each row corresponds to a token embedding. We compute prediction logits using temperature-scaled cosine similarity. Both the digit state and token embeddings are L2-normalized:
\begin{equation}
\hat{\mathbf{h}}_t^k = \frac{\mathbf{h}_t^k}{\|\mathbf{h}_t^k\|_2}, \quad
\hat{E}^k_v = \frac{E^k_v}{\|E^k_v\|_2},
\end{equation}
\begin{equation}
\boldsymbol{\ell}_t^k = \frac{\hat{\mathbf{h}}_t^k (\hat{E}^k)^\top}{\tau} \in \mathbb{R}^{V_k},
\end{equation}
where $\tau$ is the temperature hyperparameter. The probability distribution over codebook $\mathcal{W}^k$ is then:
\begin{equation}
P_\theta^k(w = v \mid \mathbf{h}_t) = \mathrm{softmax}(\boldsymbol{\ell}_t^k)_v.
\end{equation}

\section{Training Algorithm}
\label{app:algorithm}

Algorithm~\ref{alg:curriculum} summarizes the complete MHL training procedure with curriculum scheduling.

\begin{algorithm}[ht]
\caption{MHL Curriculum Training}
\label{alg:curriculum}
\begin{algorithmic}[1]
\REQUIRE Training set $\mathcal{D}$, initial mask ratio $\gamma_0$, decay factor $\eta$, minimum ratio $\gamma_{\min}$, patience $p$
\STATE Initialize model parameters $\theta$
\STATE $\gamma \leftarrow \gamma_0$, stage $\leftarrow$ \texttt{RANDOM}

\STATE \textit{// \textbf{Phase I: Random Masking Warm-up}}
\WHILE{stage $=$ \texttt{RANDOM}}
    \FOR{each batch $\mathcal{B} \in \mathcal{D}$}
        \STATE Apply random masking with ratio $\gamma$
        \STATE Compute $\mathcal{L}_{\text{MHL}} = \lambda_1 \mathcal{L}_{\text{next}} + \lambda_2 \mathcal{L}_{\text{mask}}$
        \STATE Update $\theta$ via gradient descent
    \ENDFOR
    \IF{validation loss converges for $p$ epochs}
        \STATE stage $\leftarrow$ \texttt{ENTROPY}
    \ENDIF
\ENDWHILE

\STATE \textit{// \textbf{Phase II: Entropy-Guided Masking with Decay}}
\WHILE{stage $=$ \texttt{ENTROPY} \AND $\gamma > 0$}
    \FOR{each batch $\mathcal{B} \in \mathcal{D}$}
        \STATE Compute entropy $\mathcal{H}_t^k$ via gradient-free forward pass
        \STATE Mask top-$N$ highest-entropy positions
        \STATE Compute $\mathcal{L}_{\text{MHL}}$ and update $\theta$
    \ENDFOR
    \IF{validation loss plateaus for $p$ epochs}
        \STATE $\gamma \leftarrow \max(\gamma_{\min}, \gamma \cdot \eta)$
    \ENDIF
    \IF{$\gamma \leq \gamma_{\min}$}
        \STATE $\gamma \leftarrow 0$, stage $\leftarrow$ \texttt{PREDICT}
    \ENDIF
\ENDWHILE

\STATE \textit{// \textbf{Phase III: Inference Alignment}}
\WHILE{not converged}
    \FOR{each batch $\mathcal{B} \in \mathcal{D}$}
        \STATE Compute $\mathcal{L}_{\text{next}}$ only (no masking)
        \STATE Update $\theta$ via gradient descent
    \ENDFOR
\ENDWHILE

\RETURN Trained model parameters $\theta$
\end{algorithmic}
\end{algorithm}

\section{Notation Summary}
\label{app:notation}

Table~\ref{tab:notation} summarizes the key notations used throughout this paper.

\begin{table}[ht]
\centering
\small
\begin{tabular}{cl}
\toprule
\textbf{Symbol} & \textbf{Description} \\
\midrule
\multicolumn{2}{l}{\textit{Item Representation}} \\
$\mathcal{I}$ & Item set \\
$i$ & An item in $\mathcal{I}$ \\
$K$ & Number of codebooks (digits per item) \\
$\mathcal{W}^k$ & The $k$-th codebook vocabulary \\
$V_k$ & Size of the $k$-th codebook, $|\mathcal{W}^k|$ \\
$\phi(i)$ & Semantic ID of item $i$ \\
$w_i^k$ & The $k$-th digit of item $i$'s semantic ID \\
\midrule
\multicolumn{2}{l}{\textit{Vocabulary}} \\
$\mathcal{V}$ & Unified token vocabulary \\
$\mathcal{V}_{\text{code}}, \mathcal{V}_{\text{mask}}$ & Codebook tokens and mask tokens \\
$T_{\max}$ & Maximum sequence length \\
\midrule
\multicolumn{2}{l}{\textit{Sequence and Embeddings}} \\
$S_T$ & User history sequence of length $T$ \\
$T$ & Sequence length \\
$t$ & Position index in sequence \\
$d$ & Embedding dimension \\
$e_t^k$ & Token embedding for position $t$, digit $k$ \\
$H_t$ & Item-level representation (mean-pooled) \\
\midrule
\multicolumn{2}{l}{\textit{Model Components}} \\
$\theta$ & Model parameters \\
$\mathbf{h}_t$ & Contextualized hidden state from decoder \\
$\mathbf{h}_t^k$ & Digit-wise hidden state from head $g_k$ \\
$g_k$ & The $k$-th prediction head \\
$E^k$ & Embedding matrix for codebook $k$ \\
$\tau$ & Temperature for cosine classifier \\
$\boldsymbol{\ell}_t^k$ & Prediction logits for position $t$, digit $k$ \\
$P_\theta^k$ & Prediction distribution over $\mathcal{W}^k$ \\
\midrule
\multicolumn{2}{l}{\textit{Training Objectives}} \\
$\mathcal{M}$ & Set of masked positions \\
$\mathcal{L}_{\text{next}}$ & Next-item prediction loss \\
$\mathcal{L}_{\text{mask}}$ & Masked reconstruction loss \\
$\mathcal{L}_{\text{MHL}}$ & Combined MHL loss \\
$\lambda_1, \lambda_2$ & Loss balancing weights \\
\midrule
\multicolumn{2}{l}{\textit{Entropy-Guided Masking}} \\
$\mathcal{H}_t^k$ & Entropy at position $t$, digit $k$ \\
$\bar{\mathcal{H}}_t$ & Item-level entropy (averaged) \\
\midrule
\multicolumn{2}{l}{\textit{Curriculum Training}} \\
$\gamma$ & Current masking ratio \\
$\gamma_0$ & Initial masking ratio \\
$\gamma_{\min}$ & Minimum masking ratio threshold \\
$\eta$ & Decay factor for masking ratio \\
$N$ & Number of positions to mask (budget) \\
$p$ & Patience for early stopping \\
\bottomrule
\end{tabular}
\caption{Summary of notations.}
\label{tab:notation}
\end{table}

\section{Pilot Experiment}
\label{appdix:pilot}
To illustrate the motivation of this paper, we conduct a pilot experiment to examine whether models trained with the standard next-item prediction paradigm overly rely on recent interactions, potentially neglecting the user's past behaviors. Specifically, on the \textit{Toys and Games} dataset, for sequences longer than 20 in the test set, we truncate each sequence by removing the last 15 items. For example, given an original sequence $[i_1, i_2, \ldots, i_{25}] \rightarrow i_{26}$, the truncated version becomes $[i_1, i_2, \ldots, i_{10}] \rightarrow i_{11}$. This setup evaluates how well models capture long-range dependencies when recent interactions are unavailable, rather than relying on short-term patterns near the target item.

The results of full-sequence and truncated-sequence evaluation are shown in Table~\ref{tab:sport_generalization}. Under the full sequence setting, MHL outperforms RPG across all metrics, achieving an 18.23\% improvement on N@10. In the truncated setting, which emphasizes longer-range dependencies, the improvement is even larger, reaching 43.95\%, indicating that MHL not only captures both short-term and long-term user preferences, but also better understands the overall sequence context. This comparison demonstrates that MHL more effectively models user behavior, whereas RPG tends to rely more heavily on recent interactions.

\begin{figure}[t] 
 \centering \includegraphics[width=0.8 \columnwidth]{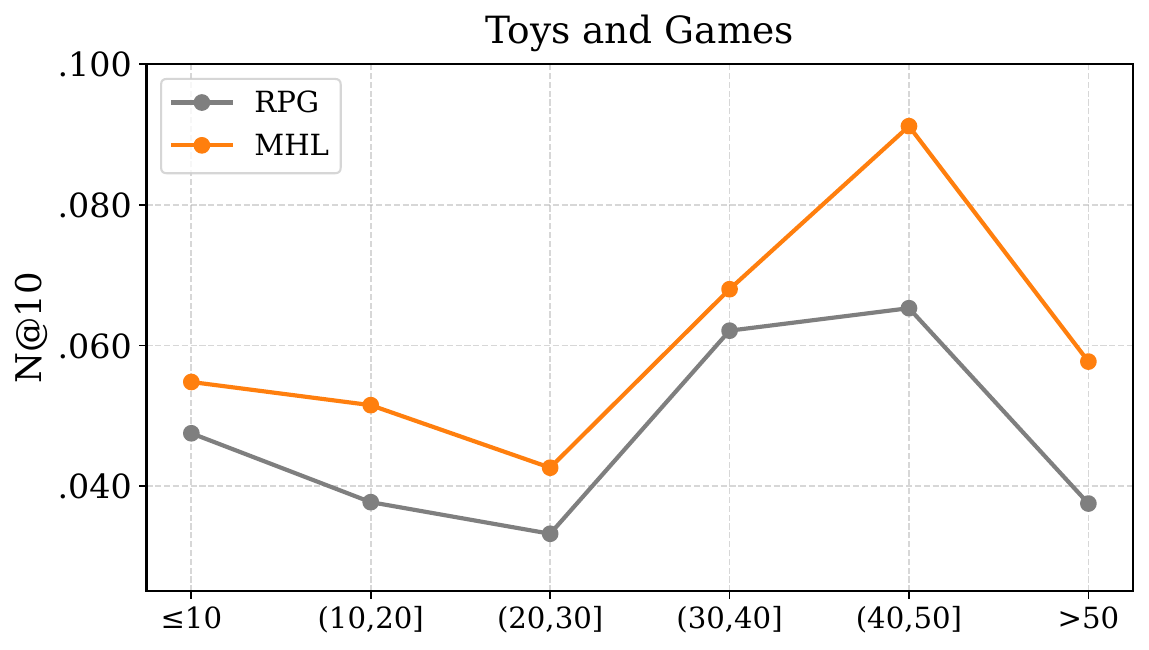} 
 \caption{Length-stratified N@10 comparison on Toys and Games. MHL adopts a 16-bit codebook consistent with RPG, and employs a mask ratio of 0.10. MHL outperforms RPG across all sequence length bins.} 
 \label{fig:pilot_test-set-length} 
\end{figure}

To further validate MHL's ability to capture long-term user intent, we conduct a length-stratified analysis. We bucket the test sequences by length and compute N@10 for both RPG and MHL. The performance trends are visualized in Figure~\ref{fig:pilot_test-set-length}, with detailed numerical results reported in Table~\ref{tab:length_stratified_n10}. RPG performs reasonably well on medium-length sequences (30–50 items) but struggles on very short and very long sequences. For instance, sequences longer than 50 items see RPG's N@10 drop to 0.0375, while MHL boosts it to 0.0577, corresponding to a 53.33\% relative improvement. Overall, MHL consistently outperforms RPG across all length buckets, and the relative improvement is most pronounced for the extremely long sequences. These results confirm that MHL effectively captures long-term user preferences rather than relying primarily on recent interactions, further supporting the motivation for our proposed approach.

\section{Hyperparameter Sensitivity Analysis}

\subsection{Codebook Size}
\label{appdix:Detailed Performance Results Across Codebook Sizes}

We investigate the influence of codebook size on model efficacy by varying the size within the range of 4 to 64. Performance exhibits a consistent upward trend as the codebook size expands from 4 to 32 across all three masking strategies and datasets. Notably, codebook sizes of 16 and 32 yield the optimal results; specifically, size 16 proves marginally superior on the Beauty dataset, whereas 32 dominates on Toys and Sports. Conversely, a size of 4 leads to significant performance deterioration due to inadequate semantic granularity, while size 64 incurs a slight decline, likely stemming from sparsity issues and increased optimization difficulty caused by excessive capacity.

These results suggest that codebook size should balance expressiveness and learnability. Based on these findings, we adopt codebook sizes of 16 for Beauty and 32 for Toys and Sports in our main experiments. Table~\ref{tab:codebook_vertical_full} provides a detailed performance comparison across different codebook sizes with a mask ratio of 0.15, including results for three mask strategies (Token-level, Item-level, and Mixed-level) on all three datasets.

\begin{table*}[ht]
\centering
\scriptsize
\setlength{\tabcolsep}{3pt} 
\resizebox{\textwidth}{!}{ 
\begin{tabular}{>{\centering\arraybackslash}p{1.5cm} >{\centering\arraybackslash}p{1.0cm} cccc cccc cccc}
\toprule
\multirow{2}{*}{\makecell[c]{\rule{0pt}{2.5ex}Mask\\Strategy}} & \multirow{2}{*}{\makecell[c]{\rule{0pt}{2.5ex}Codebook\\Size}} 
 & \multicolumn{4}{c}{Beauty} 
 & \multicolumn{4}{c}{Toys and Games} 
 & \multicolumn{4}{c}{Sports and Outdoors} \\
\cmidrule(lr){3-6} \cmidrule(lr){7-10} \cmidrule(lr){11-14}
 &  & R@5 & N@5 & R@10 & N@10 
    & R@5 & N@5 & R@10 & N@10 
    & R@5 & N@5 & R@10 & N@10 \\
\midrule
\multirow{5}{*}{\parbox[t]{1.5cm}{\raggedright Token-level}}
 & ~~4  & .0406 & .0291 & .0568 & .0343 & .0434 & .0298 & .0638 & .0363 & .0177 & .0120 & .0277 & .0152 \\
 & ~~8  & .0502 & .0369 & .0691 & .0430 & .0577 & .0409 & .0805 & .0482 & .0280 & .0209 & .0395 & .0245 \\
 & 16   & \textbf{.0574} & \textbf{.0424} & .0795 & \textbf{.0495} & \textbf{.0644} & \underline{.0456} & \textbf{.0883} & \underline{.0533} & \underline{.0334} & \underline{.0239} & .0474 & \underline{.0284} \\
 & 32   & \underline{.0568} & \underline{.0411} & \textbf{.0814} & \underline{.0490} & \underline{.0634} & \textbf{.0458} & \underline{.0880} & \textbf{.0538} & \textbf{.0344} & \textbf{.0244} & \textbf{.0495} & \textbf{.0292} \\
 & 64   & .0552 & .0390 & \underline{.0805} & .0472 & .0618 & .0448 & .0861 & .0527 & .0325 & .0226 & \underline{.0475} & .0275 \\
\midrule
\multirow{5}{*}{\parbox[t]{1.5cm}{\raggedright Item-level}}
 & ~~4  & .0344 & .0253 & .0516 & .0308 & .0385 & .0259 & .0576 & .0320 & .0151 & .0107 & .0242 & .0136 \\
 & ~~8  & .0458 & .0329 & .0646 & .0390 & .0503 & .0359 & .0716 & .0427 & .0249 & .0180 & .0366 & .0218 \\
 & 16   & .0501 & \underline{.0363} & .0704 & \underline{.0429} & \underline{.0553} & \underline{.0390} & \underline{.0802} & \underline{.0470} & \underline{.0282} & \textbf{.0197} & \underline{.0419} & \textbf{.0241} \\
 & 32   & \textbf{.0523} & \textbf{.0368} & \textbf{.0743} & \textbf{.0439} & \textbf{.0579} & \textbf{.0416} & \textbf{.0811} & \textbf{.0491} & \textbf{.0286} & \underline{.0193} & \textbf{.0421} & \underline{.0236} \\
 & 64   & \underline{.0509} & .0358 & \underline{.0731} & \underline{.0429} & .0546 & .0384 & .0768 & .0456 & .0273 & .0179 & .0417 & .0226 \\
\midrule
\multirow{5}{*}{\parbox[t]{1.5cm}{\raggedright Mixed-level}}
 & ~~4  & .0376 & .0272 & .0531 & .0322 & .0422 & .0287 & .0603 & .0346 & .0203 & .0146 & .0302 & .0178 \\
 & ~~8  & .0468 & .0343 & .0645 & .0400 & .0553 & .0385 & .0785 & .0460 & .0279 & .0201 & .0389 & .0237 \\
 & 16   & \textbf{.0537} & \textbf{.0384} & .0757 & \underline{.0455} & \underline{.0592} & \underline{.0421} & \underline{.0845} & \underline{.0503} & \underline{.0325} & \underline{.0228} & \underline{.0478} & \underline{.0278} \\
 & 32   & \textbf{.0537} & \underline{.0383} & \textbf{.0784} & \textbf{.0463} & \textbf{.0613} & \textbf{.0434} & \textbf{.0855} & \textbf{.0511} & \textbf{.0334} & \textbf{.0235} & \textbf{.0489} & \textbf{.0285} \\
 & 64   & .0533 & .0377 & \underline{.0760} & .0451 & .0579 & .0411 & .0824 & .0490 & .0318 & .0218 & .0471 & .0267 \\
\bottomrule
\end{tabular}
}
\caption{Performance comparison across different codebook sizes with mask ratio 0.15.}
\label{tab:codebook_vertical_full}
\end{table*}

\subsection{Mask Ratio}
\label{appdix:Mask Ratio Sensitivity Analysis}

We investigate the impact of the masking ratio on model robustness by varying it from 0.05 to 0.40. MHL maintains relatively stable performance across three masking strategies (Token-level, Item-level, and Mixed-level) and all datasets (Beauty, Toys, and Sports), utilizing dataset-specific codebook sizes. Notably, while the model exhibits slight performance fluctuations depending on the specific strategy and dataset, it generally demonstrates strong resilience to ratio variations.

For token-level masking, lower ratios between 0.05 and 0.15 yield consistently strong performance, with 0.05 achieving optimal results on Toys and Games and 0.10--0.15 performing best on Beauty and Sports and Outdoors. Item-level masking shows greater variability across datasets, with optimal ratios ranging from 0.10 to 0.35 depending on the specific dataset. Mixed-level masking demonstrates stable performance within the 0.10--0.20 range, often achieving competitive results across all datasets. While mixed-level masking provides a balanced and robust solution, token-level masking with lower ratios tends to achieve the highest overall performance, particularly on datasets with richer textual information.

The observed stability across mask ratios indicates that MHL is robust to variations in masking, making it suitable for real-world deployment where flexibility in masking strategies is important. Table~\ref{tab:mask_ratio_comp} presents detailed performance across different mask ratios for all masking strategies and datasets.

\begin{table*}[ht]
\centering
\scriptsize
\setlength{\tabcolsep}{3pt}
\resizebox{\textwidth}{!}{%
\begin{tabular}{>{\centering\arraybackslash}p{1.5cm} >{\centering\arraybackslash}p{1.0cm} cccc cccc cccc}
\toprule
\multirow{2}{*}{\makecell[c]{\rule{0pt}{2.5ex}Mask\\Strategy}} & \multirow{2}{*}{\makecell[c]{\rule{0pt}{2.5ex}Mask\\Ratio}} 
 & \multicolumn{4}{c}{Beauty} 
 & \multicolumn{4}{c}{Toys and Games} 
 & \multicolumn{4}{c}{Sports and Outdoors} \\
\cmidrule(lr){3-6} \cmidrule(lr){7-10} \cmidrule(lr){11-14}
 &  & R@5 & N@5 & R@10 & N@10 
    & R@5 & N@5 & R@10 & N@10 
    & R@5 & N@5 & R@10 & N@10 \\
\midrule
\multirow{8}{*}{\parbox[t]{1.5cm}{\raggedright Token-level}}
 & 0.05 & .0562 & .0410 & .0782 & .0481 & \textbf{.0657} & \textbf{.0476} & \textbf{.0898} & \textbf{.0553} & .0341 & .0236 & \textbf{.0499} & .0287 \\
 & 0.10 & \underline{.0572} & \underline{.0412} & \textbf{.0805} & \underline{.0487} & \underline{.0632} & \underline{.0463} & \underline{.0885} & \underline{.0544} & \textbf{.0346} & \textbf{.0246} & .0493 & \textbf{.0293} \\
 & 0.15 & \textbf{.0574} & \textbf{.0424} & \underline{.0795} & \textbf{.0495} & .0634 & .0458 & .0880 & .0538 & \underline{.0344} & \underline{.0244} & \underline{.0495} & \underline{.0292} \\
 & 0.20 & .0549 & .0403 & .0788 & .0480 & .0628 & .0457 & .0858 & .0531 & .0339 & .0240 & .0473 & .0283 \\
 & 0.25 & .0563 & .0409 & \underline{.0795} & .0484 & .0624 & .0447 & .0873 & .0527 & .0339 & .0239 & .0467 & .0280 \\
 & 0.30 & .0558 & .0407 & .0774 & .0476 & .0613 & .0443 & .0854 & .0521 & .0333 & .0238 & .0466 & .0281 \\
 & 0.35 & .0532 & .0389 & .0767 & .0465 & .0615 & .0439 & .0841 & .0512 & .0333 & .0236 & .0472 & .0280 \\
 & 0.40 & .0551 & .0399 & .0772 & .0471 & .0593 & .0429 & .0844 & .0510 & .0336 & .0237 & .0474 & .0282 \\
\midrule
\multirow{8}{*}{\parbox[t]{1.5cm}{\raggedright Item-level}}
 & 0.05 & .0327 & .0225 & .0499 & .0280 & .0565 & .0400 & .0778 & .0469 & .0277 & .0183 & .0418 & .0229 \\
 & 0.10 & .0495 & .0359 & .0700 & .0424 & \textbf{.0583} & \textbf{.0420} & .0810 & \textbf{.0492} & \textbf{.0301} & \textbf{.0201} & .0433 & .0244 \\
 & 0.15 & .0501 & .0363 & .0704 & .0429 & .0553 & .0384 & .0755 & .0449 & .0286 & .0193 & .0421 & .0236 \\
 & 0.20 & .0507 & \underline{.0372} & .0716 & .0439 & .0568 & \underline{.0408} & \textbf{.0819} & \underline{.0489} & .0290 & .0195 & .0426 & .0239 \\
 & 0.25 & .0511 & \underline{.0372} & .0717 & .0439 & .0568 & .0403 & .0800 & .0477 & .0292 & .0193 & .0439 & .0241 \\
 & 0.30 & .0510 & .0370 & .0715 & .0436 & \underline{.0580} & .0404 & \underline{.0815} & .0479 & .0295 & \underline{.0200} & \textbf{.0447} & \textbf{.0248} \\
 & 0.35 & \textbf{.0521} & \textbf{.0379} & \textbf{.0749} & \textbf{.0453} & .0571 & .0395 & .0807 & .0471 & .0285 & .0192 & \underline{.0446} & .0243 \\
 & 0.40 & \underline{.0516} & .0370 & \underline{.0740} & \underline{.0442} & .0555 & .0384 & .0806 & .0466 & \underline{.0296} & .0199 & .0439 & \underline{.0245} \\

\midrule
\multirow{8}{*}{\parbox[t]{1.5cm}{\raggedright Mixed-level}}
 & 0.05 & .0544 & .0392 & .0758 & .0461 & .0596 & .0426 & .0821 & .0498 & .0323 & .0225 & .0485 & .0277 \\
 & 0.10 & \underline{.0548} & \underline{.0397} & \underline{.0767} & \underline{.0467} & .0610 & \textbf{.0440} & \underline{.0855} & \textbf{.0519} & \textbf{.0342} & .0236 & .0483 & .0280 \\
 & 0.15 & .0537 & .0384 & .0757 & .0455 & .0613 & .0434 & \underline{.0855} & .0511 & .0334 & .0235 & .0489 & \underline{.0285} \\
 & 0.20 & \textbf{.0560} & \textbf{.0401} & \textbf{.0775} & \textbf{.0470} & \textbf{.0617} & \textbf{.0440} & .0853 & \underline{.0516} & \textbf{.0342} & \textbf{.0238} & \textbf{.0490} & \textbf{.0286} \\
 & 0.25 & .0529 & .0382 & .0746 & .0452 & \textbf{.0617} & .0433 & \textbf{.0860} & .0512 & .0335 & .0235 & .0480 & .0282 \\
 & 0.30 & .0519 & .0377 & .0743 & .0449 & .0612 & .0428 & .0854 & .0506 & .0335 & .0232 & .0489 & .0282 \\
 & 0.35 & .0517 & .0372 & .0757 & .0449 & .0598 & .0422 & .0848 & .0503 & .0340 & \underline{.0237} & \textbf{.0490} & \underline{.0285} \\
 & 0.40 & .0516 & .0368 & .0753 & .0444 & .0590 & .0418 & .0825 & .0494 & .0329 & .0228 & \textbf{.0490} & .0280 \\

\bottomrule
\end{tabular}%
}
\caption{Performance Sensitivity Analysis across Different Mask Ratios with Dataset-Specific Codebook Sizes (16 for Beauty, 32 for Toys and Games / Sports and Outdoors).}
\label{tab:mask_ratio_comp}
\end{table*}

\subsection{Reconstruction Loss}
\label{appdix:Recon Loss Sensitivity Analysis}

To investigate the contribution of the reconstruction objective in MHL, we vary the reconstruction loss weight from 0.2 to 2.0 while keeping the prediction loss fixed at 1.0. We observe that the optimal reconstruction weight is dataset-dependent: Beauty achieves best performance at 1.0, Toys and Games peaks at 1.8, and Sports and Outdoors performs optimally around 1.2--1.4. Despite this variation, the best results consistently fall within the range of 1.0--1.8, suggesting that a moderate-to-high reconstruction weight is generally beneficial.

Performance degrades noticeably when the weight drops below 0.8, indicating that insufficient reconstruction supervision weakens the learned token representations. Conversely, excessively high weights (e.g., 2.0) do not yield further improvements and may even hurt performance, likely due to the reconstruction objective overshadowing the prediction task. These findings highlight that reconstruction loss serves as an effective auxiliary signal that complements the primary prediction objective, but requires careful balancing to maximize performance. Table~\ref{tab:recon_loss_values} reports the specific performance metrics across all reconstruction loss values.

\begin{table*}[ht]
\centering
\resizebox{\textwidth}{!}{%
\begin{tabular}{>{\centering\arraybackslash}p{2.5cm} 
                >{\centering\arraybackslash}p{2.5cm}
                cccc cccc cccc}
\toprule
\multirow{2}{*}{\makecell[c]{\rule{0pt}{2.5ex}Mask\\Strategy}} 
& \multirow{2}{*}{\makecell[c]{\rule{0pt}{2.5ex}Reconstruction\\Loss}} 
& \multicolumn{4}{c}{Beauty} 
& \multicolumn{4}{c}{Toys and Games} 
& \multicolumn{4}{c}{Sports and Outdoors} \\
\cmidrule(lr){3-6} \cmidrule(lr){7-10} \cmidrule(lr){11-14}
&  & R@5 & N@5 & R@10 & N@10 
   & R@5 & N@5 & R@10 & N@10
   & R@5 & N@5 & R@10 & N@10 \\
\midrule
\multirow{10}{*}{\parbox[t]{2.5cm}{\centering Token-level}} 
 & 2.0 & .0562 & .0414 & .0779 & .0484 & .0638 & .0475 & .0868 & .0548 & .0337 & .0241 & .0472 & .0285 \\
 & 1.8 & .0560 & .0411 & .0769 & .0479 & \textbf{.0672} & \textbf{.0489} & \textbf{.0903} & \textbf{.0564} & .0347 & .0246 & .0484 & .0290 \\
 & 1.6 & \underline{.0564} & \underline{.0415} & .0767 & .0480 & .0653 & .0476 & .0877 & .0548 & .0344 & .0241 & .0484 & .0286 \\
 & 1.4 & .0559 & .0413 & \textbf{.0800} & \underline{.0490} & \underline{.0664} & \textbf{.0489} & .0896 & \underline{.0563} & \textbf{.0361} & \textbf{.0253} & \underline{.0503} & \textbf{.0298} \\
 & 1.2 & .0559 & .0412 & .0774 & .0481 & .0646 & .0474 & .0867 & .0545 & \underline{.0359} & \underline{.0249} & \textbf{.0511} & \textbf{.0298} \\
 & 1.0 & \textbf{.0574} & \textbf{.0424} & \underline{.0795} & \textbf{.0495} & .0657 & .0476 & \underline{.0898} & .0553 & .0346 & .0246 & .0493 & .0293 \\
 & 0.8 & .0554 & .0404 & .0793 & .0481 & .0642 & .0458 & .0890 & .0538 & .0348 & .0244 & .0482 & .0287 \\
 & 0.6 & .0549 & .0401 & .0790 & .0479 & .0640 & .0460 & .0881 & .0537 & .0331 & .0235 & .0478 & .0283 \\
 & 0.4 & .0533 & .0382 & .0768 & .0458 & .0611 & .0435 & .0856 & .0513 & .0330 & .0235 & .0465 & .0278 \\
 & 0.2 & .0526 & .0376 & .0758 & .0450 & .0602 & .0429 & .0854 & .0510 & .0330 & .0224 & .0487 & .0275 \\
\bottomrule
\end{tabular}
}
\caption{Token-level performance under varying reconstruction loss values (predict loss fixed at 1.0).  
Mask ratios: 0.15 for Beauty, 0.05 for Toys, 0.10 for Sports; codebook sizes: 16 for Beauty and 32 for Toys/Sports.}
\label{tab:recon_loss_values}
\end{table*}

\section{Cold-Start Recommendation Analysis}
\label{appdix:Cold-Start Recommendation Analysis}

\begin{table*}[ht]
\centering
\setlength{\abovecaptionskip}{0.1cm}  
\setlength{\belowcaptionskip}{-0.4cm}
\small 
\setlength{\tabcolsep}{4pt} 
\resizebox{1.0\textwidth}{!}{
\begin{tabular}{ll cccc cccc cccc}
\toprule
\multirow{2}{*}{Model} & \multirow{2}{*}{\makecell[c]{\rule{0pt}{2.5ex}Frequency\\Bin}} 
 & \multicolumn{4}{c}{Beauty} 
 & \multicolumn{4}{c}{Toys and Games} 
 & \multicolumn{4}{c}{Sports and Outdoors} \\
\cmidrule(lr){3-6} \cmidrule(lr){7-10} \cmidrule(lr){11-14}
 & & R@5 & N@5 & R@10 & N@10 
 & R@5 & N@5 & R@10 & N@10 
 & R@5 & N@5 & R@10 & N@10 \\
\midrule
\multirow{4}{*}{MHL} 
 & [0, 5]   & .0215 & .0139 & .0332 & .0177 & .0355 & .0244 & .0546 & .0306 & .0066 & .0036 & .0118 & .0053 \\
 & [6, 10]  & .0385 & .0276 & .0505 & .0314 & .0656 & .0460 & .0855 & .0524 & .0200 & .0145 & .0304 & .0179 \\
 & [11, 15] & .0528 & .0408 & .0776 & .0487 & .0863 & .0631 & .1123 & .0714 & .0271 & .0197 & .0409 & .0241 \\
 & [16, 20] & .0399 & .0302 & .0608 & .0368 & .0757 & .0555 & .1087 & .0661 & .0271 & .0194 & .0398 & .0235 \\
\midrule
\multirow{4}{*}{RPG} 
 & [0, 5]   & .0056 & .0037 & .0165 & .0071 & .0246 & .0160 & .0418 & .0215 & .0027 & .0014 & .0072 & .0028 \\
 & [6, 10]  & .0141 & .0092 & .0263 & .0131 & .0446 & .0291 & .0661 & .0359 & .0101 & .0062 & .0204 & .0095 \\
 & [11, 15] & .0390 & .0300 & .0544 & .0349 & .0678 & .0459 & .1011 & .0566 & .0185 & .0121 & .0306 & .0159 \\
 & [16, 20] & .0353 & .0254 & .0595 & .0331 & .0591 & .0426 & .0860 & .0512 & .0139 & .0094 & .0275 & .0138 \\
\bottomrule
\end{tabular}%
}
\caption{Cold-start evaluation results across different target item frequency bins. The target items are grouped by their occurrence frequency in the training set.}
\label{tab:cold_start_full}
\end{table*}

Table~\ref{tab:cold_start_full} reports the complete cold-start evaluation results across different target item frequency bins on three datasets. The results demonstrate that MHL consistently outperforms RPG across all frequency ranges, with the performance advantage being most pronounced for the coldest items. On Beauty, MHL achieves an N@10 of 0.0177 for items with training frequency [0, 5], compared to RPG's 0.0071, representing a relative improvement of over 2.5$\times$. Similar substantial gains are observed on Toys (0.0306 vs. 0.0215) and Sports (0.0053 vs. 0.0028).

Both models exhibit a consistent inverted-U trend across all datasets: performance improves from the [0, 5] bin to [6, 10] and continues rising to [11, 15], after which it declines at [16, 20]. This pattern suggests that extremely rare items lack sufficient collaborative signals for accurate prediction, while moderately frequent items ([11, 15]) are easiest to recommend due to balanced signal availability without excessive popularity competition. The decline at [16, 20] may indicate that items in this range suffer from popularity bias or increased competition from other frequent items.

Importantly, MHL's advantage persists across all frequency ranges, not just cold items. On the Toys dataset, MHL achieves N@10 scores of 0.0524, 0.0714, and 0.0661 for bins [6, 10], [11, 15], and [16, 20] respectively, consistently surpassing RPG's corresponding scores of 0.0359, 0.0566, and 0.0512. This validates that MHL's masked history learning provides robust representations that generalize well regardless of item popularity, enabling the model to effectively leverage limited historical signals for cold items while maintaining accuracy for more frequent items.

\section{Statistics of the Dataset}
\label{appendix:statistic}

The detailed statistics of Amazon Reviews 2014 datasets is shown in Table~\ref{tab:dataset_stats}.

\begin{table}[t]
\centering
\scriptsize
\setlength{\tabcolsep}{3pt}
\resizebox{\columnwidth}{!}{
\begin{tabular}{>{\centering\arraybackslash}p{1.8cm} >{\centering\arraybackslash}p{1.8cm} cccc c}
\toprule
\multirow{2}{*}{\makecell[c]{\rule{0pt}{2.5ex}Setting}} & 
\multirow{2}{*}{\makecell[c]{\rule{0pt}{2.5ex}Model}} 
 & \multicolumn{4}{c}{Toys and Games} 
 & \multirow{2}{*}{\makecell[c]{\rule{0pt}{2.5ex}$\Delta$\%}} \\
\cmidrule(lr){3-6}
 &  & R@5 & N@5 & R@10 & N@10 &  \\
\midrule
Full & RPG
 & .0550 & .0386 & .0778 & .0460 & -- \\
Full & MHL (ours)
 & .0656 & .0471 & .0885 & .0544 & \textbf{+18.3} \\
\midrule
Truncated & RPG
 & .6355 & .4823 & .7454 & .5176 & -- \\
Truncated & MHL (ours)
 & .8460 & .7208 & .9199 & .7451 & \textbf{+44.0} \\
\bottomrule
\end{tabular}
}
\caption{Comparison of RPG and MHL on the \textit{Toys} dataset. Both RPG and MHL use a 16-bit codebook, with MHL utilizing a mask ratio of 0.10. ``Full'' denotes evaluation on complete sequences, while ``Truncated'' denotes evaluation on the prefixes of long sequences. $\Delta$\% denotes the relative improvement of N@10.}
\label{tab:sport_generalization}
\end{table}

\begin{table}[t]
\centering
\scriptsize
\setlength{\tabcolsep}{6pt}
\resizebox{\columnwidth}{!}{
\begin{tabular}{>{\centering\arraybackslash}p{2.2cm} >{\centering\arraybackslash}p{1.8cm} >{\centering\arraybackslash}p{1.8cm} >{\centering\arraybackslash}p{1.5cm}}
\toprule
\multirow{2}{*}{\makecell[c]{Test Set}} & \multicolumn{2}{c}{Toys and Games} & \multirow{2}{*}{\makecell[c]{$\Delta$\%}} \\
\cmidrule(lr){2-3}
 & RPG N@10 & MHL N@10 &  \\
\midrule
Full       & .0460 & .0544 & \textbf{+18.3} \\
$\leq 10$  & .0475 & .0548 & \textbf{+15.4} \\
(10,20]    & .0377 & .0515 & \textbf{+36.6} \\
(20,30]    & .0332 & .0426 & \textbf{+28.3} \\
(30,40]    & .0621 & .0680 & \textbf{+9.5} \\
(40,50]    & .0653 & .0912 & \textbf{+39.7} \\
$>50$      & .0375 & .0577 & \textbf{+53.9} \\
\bottomrule
\end{tabular}
}
\caption{Length-stratified N@10 performance of RPG and MHL on the \textit{Toys and Games} test set. Both RPG and MHL use a 16-bit codebook, and MHL employs a mask ratio of 0.10. $\Delta$\% denotes the relative improvement of MHL over RPG.}
\label{tab:length_stratified_n10}
\end{table}

\begin{table}[ht]
\centering
\resizebox{\columnwidth}{!}{ 
\begin{tabular}{lcccc}
\toprule
Datasets & \#Users & \#Items & \#Interactions & Avg. $t$ \\
\midrule
Beauty & 22,363 & 12,101 & 176,139 & 8.87 \\
Toys and Games & 19,412 & 11,924 & 148,185 & 8.63 \\
Sports and Outdoors & 18,357 & 35,598 & 260,739 & 8.32 \\
\bottomrule
\end{tabular}
}
\caption{Statistics of the Amazon Reviews 2014 datasets. ``Avg. $t$'' denotes the average number of interactions per input sequence.}
\label{tab:dataset_stats}
\end{table}

\section{Implementation Details}
\label{implemental_Details}

We encode item metadata (title, brand, price, features, categories, description) using Sentence-T5 and reduce 768-dimensional embeddings to 128 dimensions with PCA. Following RPG~\citep{hou2025generating}, we discretize continuous representations into generative semantic IDs using FAISS-based OPQ. Each item is represented as a sequence of 32 tokens (32 codebooks with 256 codewords each). Our backbone is a Transformer decoder with the same parameter size as RPG~\citep{hou2025generating}: hidden size 448, 2 layers, 4 attention heads, feed-forward dimension 1024, and GELU activation. The maximum sequence length is 50 with dropout 0.3 for embeddings and attention modules. 

For training, we jointly optimize next-item prediction and masked token reconstruction with equal weights. We use entropy-guided curriculum masking: training starts with random masking, then switches to entropy-based masking; if validation does not improve for 5 consecutive evaluations, mask ratio decays linearly by $0.1\times r_0$ (with $r_0=0.15$) until reaching 0. After this, the model trains purely on prediction with early stopping patience of 20. Entropy forward propagation stabilizes masking decisions using window size 3, decay factor 2.0, and residual mixing coefficient 0.2 across item-level and token-level entropies. 

During inference, we follow RPG~\citep{hou2025generating} and apply graph-constrained beam search with beam size 50, each node keeping 50 edges, and 3 propagation steps. Optimization uses AdamW with learning rate 5e-4, batch size 64, weight decay 0.0, gradient clipping 1.0, 10k warmup steps, and cosine learning rate scheduling. We train for up to 300 epochs with early stopping patience of 20. All experiments use NVIDIA RTX A6000 GPUs with distributed training and mixed precision.

\section{Efficiency Analysis and Training Overhead}
\label{appendix:efficiency}

We conduct a controlled efficiency study on a single NVIDIA RTX A6000 GPU using the Toys and Games dataset. Both MHL and RPG use identical configurations: codebook size = 16, batch size 64, sequence length 50, AdamW optimizer with learning rate 5e-4, and mixed-precision training (FP16). As shown in Table~\ref{tab:efficiency_comparison}, MHL improves NDCG@10 by 12.6\% with an 86\% increase in step time and a modest 9.5\% memory overhead. The additional cost does not primarily originate from entropy estimation---although entropy statistics are updated dynamically at every training step ($k=1$), they are computed under {\tt torch.no\_grad()} context without gradient propagation, so entropy calculation adds negligible runtime overhead. Profiling indicates that the dominant overhead comes from the auxiliary masked reconstruction objective, which requires additional forward computation. MHL requires more epochs to converge (122 vs. 68) due to the three-phase curriculum schedule. Importantly, this overhead is confined to training; at inference time, MHL uses the same forward path as RPG with no masking or reconstruction loss, so inference latency is identical to RPG.

\begin{table*}[h]
\centering
\small
\setlength{\tabcolsep}{6pt}
\begin{tabular}{lccc}
\toprule
\textbf{Metric} & \textbf{RPG} & \textbf{MHL} & \textbf{Relative Change} \\
\midrule
Wall-clock Time & 25.4 min & 55.5 min & +118\% \\
Step Time & 64.8 ms & 120.8 ms & +86\% \\
Peak GPU Memory & 10,118 MB & 11,076 MB & +9.5\% \\
Best Epoch & 68 & 122 & +79.4\% \\
NDCG@10 & 0.0460 & 0.0518 & +12.6\% \\
\bottomrule
\end{tabular}
\caption{Efficiency comparison between RPG and MHL on Toys and Games dataset. Both models use codebook size = 16.}
\label{tab:efficiency_comparison}
\end{table*}

\end{document}

%% file: math_commands.tex
\usepackage{amsmath,amsfonts,bm}

\def\eqref#1{equation~\ref{#1}}

\def\1{\bm{1}}

\DeclareMathAlphabet{\mathsfit}{\encodingdefault}{\sfdefault}{m}{sl}
\SetMathAlphabet{\mathsfit}{bold}{\encodingdefault}{\sfdefault}{bx}{n}

